\documentclass[12pt,letterpaper]{article} 

\usepackage[english]{babel} 
\usepackage[utf8x]{inputenc} 

\usepackage{lipsum} 
\usepackage{setspace} 
\usepackage{microtype} 

\usepackage[margin=1.5in]{geometry} 
\usepackage{changepage} 

\usepackage{amsmath} 
\usepackage{amssymb} 
\usepackage{bm} 
\usepackage[retainorgcmds]{IEEEtrantools} 

\usepackage{graphicx} 
\usepackage[font=normalsize]{caption} 
\usepackage{subfig} 
\usepackage{subcaption} 

\usepackage{booktabs} 
\usepackage{tabularx} 
\usepackage{longtable} 
\usepackage{adjustbox} 
\usepackage{tabularray} 
\usepackage{multirow} 
\usepackage{makecell}    

\usepackage{listings} 
\usepackage{verbatim} 

\usepackage{xurl}
\usepackage[colorlinks=true]{hyperref} 
\hypersetup{ 
  colorlinks,
  citecolor=blue,
  linkcolor=red,
  urlcolor=cyan
  }

\usepackage[colorinlistoftodos]{todonotes} 
\usepackage{soul} 
\sethlcolor{yellow} 

\usepackage{enumitem} 
\usepackage{pifont} 
\usepackage[official]{eurosym}

\usepackage{xcolor} 
\usepackage{float} 
\usepackage{ragged2e} 
\usepackage{lineno} 
\usepackage{multicol} 
\usepackage{pdflscape} 
\usepackage{lscape} 
\usepackage[super]{nth} 

\usepackage{natbib} 

\usepackage{authblk}

\usepackage{fancyhdr} 
\fancyhfoffset[L]{0cm} 
\fancyhfoffset[R]{0cm} 
\fancypagestyle{abstractstyle}{
    \fancyhf{} 
    \fancyfoot[C]{\textcolor{gray}{\thepage}}

}

\begin{document}


\title{
\vspace{-1cm} 
\textbf{Place-Based Policy from Space: Evidence from Italy’s Inner Areas} 
}

\newcommand{\thetitle}{Blinding Lights}

\author{\href{https://orcid.org/0000-0002-9621-4008}{\includegraphics[width=0.35cm]{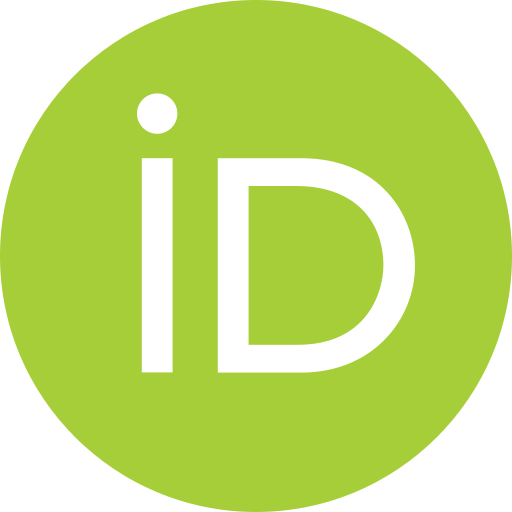}} Nicola Caravaggio$^{a,\dag,\star}$} 
\author{\href{https://orcid.org/0009-0007-7801-2449}{\includegraphics[width=0.35cm]{Figures/orcid_id.png}} Giuseppe Di Renzo$^{a}$} 
\author{\href{https://orcid.org/0000-0002-3913-0510}{\includegraphics[width=0.35cm]{Figures/orcid_id.png}} Giuliano Resce$^{a}$} 
\author{\href{https://orcid.org/0009-0004-5369-1594}{\includegraphics[width=0.35cm]{Figures/orcid_id.png}} Agapito Emanuele Santangelo$^{a}$} 
\author{\href{https://orcid.org/0000-0002-3553-0269}{\includegraphics[width=0.35cm]{Figures/orcid_id.png}} Giulia Valeria Sonzogno$^{b}$} 
\affil{\footnotesize
  $^{a}$\textit{Department of Economics, University of Molise, Campobasso -- Italy} \\
  $^{b}$\textit{Department of Economics, University of Roma Tre, Rome -- Italy} 
  \vspace{0.3cm}
  \begin{minipage}{11cm}~\\
  \centering
  \footnotesize
  $^{\dag}$\tt\href{mailto:nicola.caravaggio@unimol.it}{nicola.caravaggio@unimol.it} 
  \vspace{-1cm} 
  \end{minipage}
  }

\date{\small\today}

\renewcommand\Authands{, and } 

\maketitle


\begin{adjustwidth}{-0.25in}{-0.25in} 
\begin{abstract}

\thispagestyle{abstractstyle}

\noindent\rule[0.5ex]{\linewidth}{0.5pt}

\footnotesize

This paper evaluates a place-based policy, Italy's National Strategy for Inner Areas (SNAI), using satellite nighttime lights. Targeting structurally fragile and peripheral municipalities, SNAI invests in essential services and economic activities. Staggered difference-in-differences compares 941 funded municipalities with 3,112 eligible controls over 2011--2022. Funding is associated with an expansion of lit area and no detectable change in radiance per lit cell, suggesting that activity extends its footprint rather than intensifying in existing centres. The expansion holds across comparison groups and, descriptively, concentrates at intermediate funding per resident. Annual satellite data can complement slower and coarser statistics.

\noindent\rule[0.5ex]{\linewidth}{0.5pt}

\end{abstract}
\end{adjustwidth}

\begin{quote}

	
 
 
\end{quote}

\pagebreak


\onehalfspacing 


\section{Introduction}
\label{section:introduction}

Europe's territorial divide is increasingly framed as a metropolitan versus peripheral one, cutting across the conventional urban-rural line \citep{iammarino2019regional, ec2024cohesion9}. Peripheral areas combine structural disadvantages (poor connectivity, limited services, depopulation) with underused local assets. Reversing their decline has become a key challenge for the European Union (EU) Cohesion Policy (CP), both for its economic and social costs and for its political consequences \citep{rodriguezpose2018revenge, dijkstra2020geography}.

In response, regional development policy has shifted away from place-neutral approaches towards place-based strategies \citep{barca2012case}. By combining targeted investment with local knowledge and multilevel governance, these strategies seek to mobilise underutilised development potential and foster territorial cohesion \citep{barca_2009}. However, the evidence on the effectiveness of place-based policies remains inconclusive. On the one hand, their impacts appear heterogeneous and highly dependent on local institutional capacity and territorial context, with lagging regions often less able to mobilise their potential \citep{becker2013absorptive,beer2020every,di2025determinants,fratesi2025four,bergantino2026far}. On the other hand, this evidence largely relies on conventional indicators of employment, firms, and investment \citep{kline2014people,neumark2015place}, which are released with long lags and at coarse spatial scales, and may therefore fail to detect the early and localised activation of underused potential that these policies aim to trigger, particularly in peripheral areas.

This paper contributes to this debate by evaluating the Italian National Strategy for Inner Areas (SNAI) - one of the most comprehensive place-based programmes targeting peripheral areas in the EU - using satellite night-time radiance, which allows us to trace where economic activity emerges within a territory and to detect localised changes that may precede broader socioeconomic adjustments \citep{chen2011using,henderson2012measuring,gibson2020night}. Introduced in 2014, the SNAI targets the so-called Inner Areas (IAs), \textit{i.e.} municipalities classified by travel time to the nearest hub providing essential services (education, hospital care and rail transport) \citep{dps2014aree}, through investments in local services and productive activities aimed at counteracting depopulation and strengthening local economic activity \citep{uval2014strategia}. A growing literature has evaluated the programme, documenting heterogeneous and often localised effects on demographic trends and business dynamics \citep{benassi2024local,di2025regional,monturano2025short,bergantino2026far,bolzoni2026repopulating,mariani2026place}. However, this evidence shares the same limitation noted above.

Building on recent applications of nighttime lights to the evaluation of Cohesion Policy \citep{bachtrogler2023eu}, we move beyond the conventional approach of summarising radiance into a single municipal indicator and introduce a measure that distinguishes between two conceptually different margins of change, corresponding to distinct mechanisms through which place-based policies may operate that an aggregate radiance measure cannot separate: (i) the expansion of illuminated areas, reflecting the activation of previously underused spaces; and (ii) the intensification of radiance within already illuminated locations, reflecting the strengthening of existing economic centres (Section~\ref{subsection:measurement}).

We then combine these nighttime-light indicators with data on SNAI funding to build a municipal panel covering 2011--2022. Exploiting the staggered timing of programme implementation across municipalities, we estimate causal effects of SNAI using the multi-period difference-in-differences (DiD) estimator of \citet{callaway2021difference}. In particular, the design compares 941 municipalities that began receiving SNAI funding between 2014 and 2021 with 3,112 municipalities where funds never arrived or have not yet arrived all of them classified as IAs.

We find that SNAI funding expands the spatial footprint of economic activity (measured using nightlights as proxies) around existing settlements with no effects in intensifying established centres, consistent with the mobilisation of underused local potential. The effect is robust across comparison groups and sectors, but emerges only above a minimum funding scale and plateaus beyond a critical mass.

In doing so, this paper makes three contributions to the literature on place-based policies. First, it introduces an outcome indicator that detects local economic change earlier and at a finer spatial scale than conventional measures. Second, it proposes an evaluation methodology that overcomes a well-known limitation of official statistics, namely the lag between when a phenomenon occurs and when data on it become available for analysis. Third, it provides timely evidence on a major nationwide place-based programme targeting peripheral areas, places marked by long-term depopulation and limited access to essential services \citep{phelps2026abandoned}, at a moment when these trends continue to spread across the EU, and the future of CP is under debate \citep{ec2026proposal20282034}. In addition, we show that a documented shift in the background radiance of the widely used VIIRS annual nighttime-lights product from 2017 distorts counts of lit cells built at low thresholds, and propose an extensive-margin radiance measure based on a fixed threshold that is directly portable to other fine-scale spatial evaluations of policy interventions.

\section{Institutional Setting}
\label{section:institutional_setting}

\begin{figure}[H]
	\centering
	\caption{\textit{Italian Inner Areas and municipalities receiving resources during the SNAI 2014--2020}}
	\includegraphics[width=0.6\linewidth]{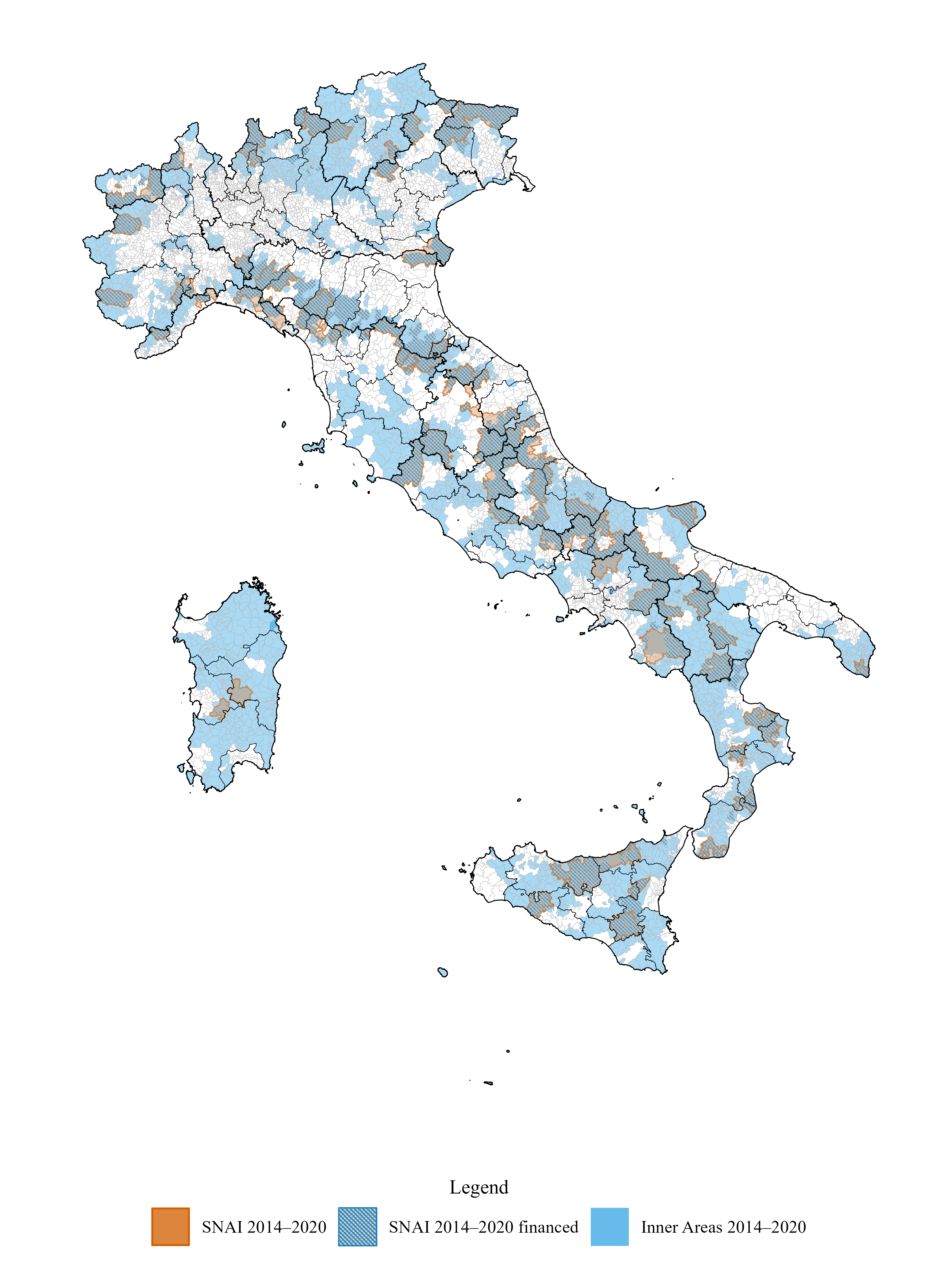}
	\label{fig:map_areas}
	\begin{minipage}{0.8\linewidth}
		\justifying
		\tiny
		\noindent
		Notes: financed municipalities for the 2014--2020 SNAI are retrieved from OpenCoesione \citep{pcm_opencoesione} (updated October 31, 2025). Municipalities are considered funded if associated with at least one active project. Inner areas include intermediate, peripheral, and ultra-peripheral municipalities. \\
		Sources: authors' elaboration based on \cite{istat_aree_interne,istat_confini,pcm_aree_snai,pcm_opencoesione}.
	\end{minipage}
\end{figure}

The SNAI was launched in 2014 \citep{uval2014strategia}, based on the introduction of a new classification which identified, for the first time, the IAs. They are territories affected by persistent social, economic, and environmental challenges, and characterised by long-term out-migration and population ageing, economic and social decline, land abandonment, and landscape transformation. The classification first identifies service-providing hubs (\textit{Poli}): municipalities offering full access to secondary education, hospital care, and rail transport, considered as essential services.\footnote{Formally, a \textit{polo} is a municipality, or a cluster of contiguous municipalities, offering at least a high school and a technical-professional institute, at least one hospital with a Level~I emergency department (a hospital with an emergency room and the capacity to handle complex emergencies, including resuscitation services), and at least one regional railway station.} Every other municipality is then classified by driving time to its nearest hub: belt (under 20 minutes), intermediate (20--40), peripheral (40--75), and ultra-peripheral (over 75 minutes) \citep{dps2014aree}. Following this classification, IAs comprise intermediate, peripheral and ultra-peripheral municipalities and account for 4,261 municipalities (53\% of the national total), in which reside about 20\% of the population, spreading over more than 60\% of the national territory (Figure \ref{fig:map_areas}).

These territories are predominantly rural and mountainous, caught in a negative spiral of depopulation and economic decline that originated during Italy's urbanisation in the 1950s and has left them increasingly vulnerable \citep{doria1958dieci,caravaggio2025aree}.

To counter these dynamics, SNAI combines two lines of intervention: strengthening essential services and supporting local development projects. These are pursued jointly to improve the welfare of the population, stimulate economic opportunities, and make better use of territorial capital, with the ultimate aim of reversing negative demographic trends \citep{uval2014strategia}.

From a methodological and theoretical standpoint, SNAI marks a departure from traditional top-down redistributive policies, acknowledging that IAs possess untapped territorial capital \citep{camagni2013regional} that cannot be activated through one-size-fits-all interventions, and relies instead on a multi-level governance model designed to elicit local knowledge through participatory planning, while conditioning central government funding on measurable outcomes and structural reforms in essential local services.

The operational implementation of the strategy follows a multi-stage selection process for local project areas (\textit{Aree Progetto}) led by a dedicated Technical Committee for Inner Areas (\textit{Comitato Tecnico Aree Interne} -- CTAI). Contiguous groups of intermediate, peripheral, and ultra-peripheral municipalities are first pre-selected using socioeconomic vulnerability criteria (\textit{e.g}., demographic conditions, migration dynamics, land abandonment, and employment levels) and then confirmed through a dialogue between regional governments, national authorities, and local stakeholders \citep{uval2014strategia, lucatelli2022}.

As a result, for the 2014--2020 period, 1,060 municipalities, grouped into 72 project areas (13.2\% of the national total and 3.5\% of the population), were selected for implementation (Figure \ref{fig:map_areas}). 
To access funding, provided through a combination of national (established through annual budget laws) and European resources,\footnote{More precisely, SNAI was supported by three European Structural and Investment (ESI) funds: the European Regional Development Fund (ERDF), the European Social Fund (ESF), and the European Agricultural Fund for Rural Development (EAFRD).} each selected area was mandated to draft a long-term strategy area (\textit{Strategia di Area}) framing local interventions within a unified theory of change that links service enhancement (education, health, and mobility) directly to localised development goals. Reflecting the policy's experimental design, interventions were governed through framework agreements (\textit{Accordi di Programma Quadro}) negotiated for each pilot area \citep{adp1420}.

Overall, under the 2014-2020 programming period, SNAI project areas received about {\euro} 800 million allocated across 3,565 projects, according to \href{https://opencoesione.gov.it/it/}{OpenCoesione}, the governmental monitoring system \citep{pcm_opencoesione}.

These projects span eleven thematic areas, from infrastructure and business support to social services (Table~\ref{tab:snai_projects_theme} for details). Their geographic distribution is shown in Figure~\ref{fig:map_areas}, which identifies funded municipalities, while Figure~\ref{fig:map_funds} provides more detail on funding intensity.\footnote{The strategy was revised for the periods 2021--2027, but monitoring of this phase is still incomplete. Our analysis therefore covers only the 2014--2020 cycle.}

\begin{figure}[H]
	\centering
	\caption{\textit{Night-light radiance for 2013 and 2022}}
    
	\noindent
	\begin{minipage}{0.49\linewidth}
		\centering
		\textbf{a}
	\end{minipage}
	\hfill
	\begin{minipage}{0.49\linewidth}
		\centering
		\textbf{b}
	\end{minipage}
    
	\includegraphics[width=\linewidth]{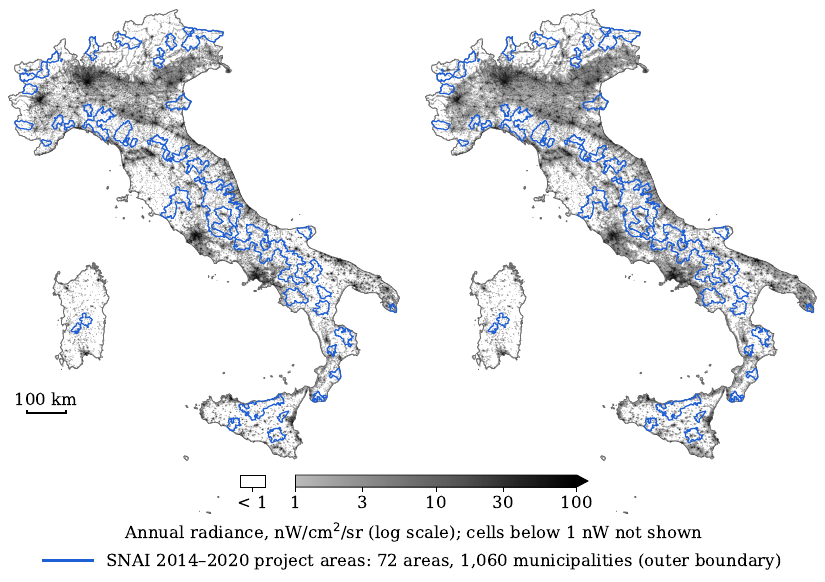}
  
	\label{fig:map_italy}
	\begin{minipage}{0.95\linewidth}
		\justifying
		\tiny
		\noindent
		Notes: annual VIIRS VNL composites, average band, for 2013 (panel a) and 2022 (panel b), drawn on the $500 \times 500$ m cells of the outcome panel in EPSG:32632 (UTM zone 32N), without reprojection. Both panels share one logarithmic colour scale. Cells below 1 nW/cm$^2$/sr are drawn as background, because below that level the background of the product changes between releases: the 2022 composite (version 2.2) sets its minimum at 0.2 nW/cm$^2$/sr, while half of the cells of the 2013 raster frame (version 2.1), open sea included, lie below 0.15. The map is masked to Italy: only cells assigned to a 2021 municipality by the centroid rule of the panel are drawn, so the lit cells shown are those counted by the lit-area outcome, 457,601 in 2013 and 571,691 in 2022 (+24.9 percent). This change includes a nationwide brightening and a single-year step of 76,628 cells between 2016 and 2017, when the background level of the product rises, about two-thirds of the total. The difference-in-differences compares treated and control municipalities in the same year and nets out both components to the extent that they are common to the two groups; the figure is therefore descriptive. Blue lines mark the outer boundary of the 72 SNAI project areas of the 2014--2020 cycle, which group 1,060 municipalities; boundaries between contiguous municipalities and between contiguous project areas are not drawn. Grey lines are regional boundaries, the coastline and the national border. \\
		Sources: authors' elaboration based on \cite{elvidge2021annual,istat_aree_interne,istat_confini,pcm_aree_snai,pcm_opencoesione}.
	\end{minipage}
\end{figure}

Figure~\ref{fig:map_italy} places the policy on the physical object we measure. Panel (a) shows the cells at or above 1 nW/cm$^2$/sr in 2013, the year before the first SNAI-funded projects started, while panel (b) shows the cells above the same threshold in 2022, the last year of our analysis; the blue lines mark the outer boundary of the 72 project areas of the 2014--2020 cycle. The radiance surface is dominated by the Po Valley, the coastal belts and the major urban systems, whereas the project areas lie mostly along the Alpine arc and the Apennine ridge, with further areas in Sicily and Sardinia, where lit cells are few and dim. Between the two years, the lit area of Italian municipalities grows from 457,601 to 571,691 cells. This growth is national in scope, and about two-thirds of it occurs in a single step between 2016 and 2017, when the background level of the product rises. 

\section{Literature review}
\label{section:literature_review}

\subsection{Evaluating place-based policies}
\label{subsection:place_based}

Place-based policies generally target lagging or structurally disadvantaged territories with interventions tailored to local conditions, on the premise that persistent regional disparities cannot be addressed by spatially neutral policies alone \citep{barca_2009, barca2012case}. Rather than compensating places for their disadvantage, they seek to address location-specific market and coordination failures through combined investment in infrastructure, public services, institutional capacity, and local productive systems.

The empirical literature on place-based policies generally documents heterogeneous effects across territories. For example, \cite{kline2014people} show that geographically targeted subsidies can generate persistent increases in local employment and economic activity, although the magnitude of these effects varies substantially across locations. Similarly, \cite{neumark2015place} find mixed evidence on the effectiveness of enterprise zones and local development programs, concluding that policy outcomes depend critically on local economic conditions, policy design, and implementation quality.

EU Cohesion Policy, the largest spatially targeted policy in Europe, offers the richest body of causal evidence, although most of it concerns programming periods that predate its explicit place-based reorientation \citep{barca_2009}. Exploiting the discontinuity in Objective 1 eligibility at 75\% of EU average GDP per capita, \cite{becker2010going} and \cite{pellegrini2013measuring} find that structural funds raise growth in recipient regions, a result confirmed within Italy using a spatial discontinuity design by \cite{giua2017spatial}. These average gains, however, conceal marked heterogeneity. They materialise only in regions with sufficient human capital and institutional quality to absorb the transfers \citep{becker2013absorptive}, and they differ substantially across member states \citep{crescenzi2020one}. Returns also depend on funding intensity: beyond a certain level, additional transfers yield no further growth \citep{becker2012too, cerqua2018spending}. Nor are the effects necessarily persistent, as growth slows once regions lose eligibility and funding is phased out \citep{barone2016boulevard}. 

Evidence on earlier national place-based instruments in Italy is similarly sobering. Territorial Pacts, bottom-up agreements among neighbouring municipalities, local entrepreneurs and trade unions, did not significantly affect employment or the number of plants, even among the pacts regarded by policymakers as best-performing \citep{accetturo2012policies}. State-funded private investment under the \textit{Contratti di Programma}, in turn, produced small gains in plants and employment in targeted municipalities, which vanished at the level of local labour markets as activity was drawn away from neighbouring areas \citep{andini2016local}.

Within this broader literature, empirical evidence on the SNAI, which represents the core of place-based turn in CP during the 2014–2020 programming period following \cite{barca_2009} in Italy, is still limited and only beginning to emerge. \cite{di2025regional} documents a positive effect on tourist nights, mainly driven by domestic tourism, while also uncovering substantial heterogeneity across geographical areas and types of expenditure. \cite{monturano2025short}, instead, find positive short-term effects on business plants and evidence of spatial spillovers to neighbouring municipalities. The work of \cite{mariani2026place} shows how the SNAI's effectiveness is highly conditional on local characteristics: positive impacts on business creation are concentrated in intermediate municipalities with low market accessibility, industrial specialisation, and strong administrative capacity, while average effects on population, employment, and housing prices remain limited. Importantly, once these local fundamentals are accounted for, the North-South divide ceases to explain policy outcomes, underscoring that institutional and structural conditions matter more than geography \textit{per se}. More generally, \cite{bergantino2026far} recalculates distances from essential services five years after the classification developed by \cite{nuvap2020strategia}, documenting a 20\% increase in the number of municipalities classified as inner areas.

Despite their growing prominence, place-based policies remain difficult to evaluate. Official statistics are often too coarse in space to capture short-term changes in local economic conditions or the heterogeneous effects these policies aim to generate. This has fuelled interest in unconventional data, such as digital traces and satellite imagery. Among these, night-time light intensity is grounded in spatial equilibrium models: changes in local productivity, accessibility and amenities affect economic activity and, in turn, artificial illumination \citep{redding2008costs,redding2017quantitative}.

Building on this literature, we use high-resolution satellite data as a timely proxy for local economic activity. Compared with existing evaluations, our approach: i) captures, at the municipal level, localised responses that administrative statistics struggle to detect \citep{gibson2021which}; ii) combines remotely sensed data with administrative records on SNAI implementation, extending the outcomes available beyond traditional socioeconomic indicators; iii) reduces the lag between economic change, its measurement and its empirical use, allowing a more timely assessment of policy impacts and their dynamics.

\subsection{Measuring economic activity from space}
\label{subsection:night_lights}

Night-time radiance entered economics as a way of measuring activity where national accounts fail in terms of granularity and timeliness. This consideration is particularly relevant in data-poor contexts: \citet{chen2011using} argued that luminosity adds information precisely where statistical systems are weakest, and \citet{henderson2012measuring} turned the intuition into a workhorse by estimating the elasticity of lights to income and using it to revise national growth figures. Later \cite{donaldson2016view} surveyed the broader use of satellite data in economics, and this proxy is now routine in spatial and development work, from the measurement of regional inequality \citep{zhou2015nighttime} to the evaluation of transport infrastructure \citep{guo2020impact}, special economic zones \citep{phommachanh2024impact}, and disaster recovery \citep{nguyennoy2020measuring}. \cite{bruederle2018nighttime} find that radiance tracks assets, education and health outcomes across African localities. \cite{mellander2015nighttime} sharpen the distinction with Swedish micro-data: lights correlate strongly with population density and establishment counts but weakly with wages and income. At the global scale, \citet{castells2021urbanising} use the lit surface itself as a description of urban form, summarising large cities by the within-city spatial Gini of night light and by its spatial autocorrelation, and report that climate-driven city growth is associated with less monocentric and more fragmented urban structures. Together, these results indicate that radiance measures the physical footprint of settlement and activity, and that the further one moves from that footprint towards welfare, the weaker the mapping becomes.

A second strand shows that results depend on the product used. \cite{gibson2020night} survey the available sources, and \cite{gibson2021which} document that estimated elasticities differ systematically between the Defense Meteorological Satellite Program (DMSP) and VIIRS generations and between urban and rural areas, so that conclusions can turn on a choice often made for convenience. The DMSP series lacks on-board calibration, and its blurring and top-coding compress the measured distribution, so that it understates spatial inequality \citep{zhang2023remotely}; harmonised DMSP-VIIRS series \citep{li2020harmonized} lengthen the panel at the cost of splicing two sensors, and \cite{gibson2021better} discusses how far the extension can be pushed. Appendices~\ref{subsubsection:appendix_products} and~\ref{subsubsection:appendix_viirs} describe the four product families, the corrections proposed for DMSP and why our design requires the VIIRS annual composites. 

A third strand asks what the errors of a product do to an estimate. \citet{jain2020benefits} distinguishes random error uncorrelated with the treatment, which increases standard errors when the satellite variable is the dependent variable, from systematic error, which can bias the coefficients of an otherwise well-identified analysis. Among the sources of systematic error, \citet{wuepper2025satellite} list the temporal inconsistencies between successive versions of a product as its producers revise the processing chain, and single out binary pixel-level outcomes, the form of our lit-cell indicator, as especially exposed, since their measurement error is necessarily negatively correlated with the truth. \citet{alixgarcia2023remotely} provide the applied case closest to ours: in the Mexican conservation programme, they evaluate whether participation is correlated with the terrain and cloud cover that affect the detection of forest loss, and modelling the misclassification limits the bias in the estimated effectiveness. The background shift described in Section~\ref{subsection:measurement} is a systematic error.

Closest to our research question is \cite{perezsindin2025can}, who ask whether radiance can proxy economic inequality at the scale of local administrative units: light-based Gini coefficients correlate only modestly with income-based ones across municipalities but track them over time, and the authors relate the divergences to infrastructure, urban design and informal activity. Light thus records more than income, including how activity is arranged in space, the margin our design isolates. Inner areas are sparsely populated, dimly lit and peripheral, the configuration in which the elasticity of lights to economic activity is weakest and the product artefacts are largest in relative terms, so we treat measurement as part of the research design (Section~\ref{subsection:measurement}). The closest application in the policy domain is \cite{bachtrogler2023eu}, who use night-light emissions to proxy local economic development in the municipalities of the border area between the Czech Republic, Germany, and Poland, and find that more EU Cohesion funding over 2007--2013 is associated with increased economic activity at the municipal level. Relative to that contribution, we separate two margins conflated in aggregate radiance (the expansion of the illuminated footprint and the brightening of existing centres), limit the background-shift artefact that would otherwise inflate the first, and characterise the response by thematic area and funding intensity.


\section{Data and Methodology}
\label{section:methodology_data}

\subsection{Night-light outcomes and a drifting background}
\label{subsection:measurement}

The outcomes are built from the annual VIIRS nighttime light (VNL) composites \citep{elvidge2021annual}, resampled to cells of $500 \times 500$ m and aggregated to municipal boundaries \citep{istat_confini}. A cell is \textit{lit} in a given year when its radiance reaches a threshold $\tau$ fixed across years. Three outcomes follow for each municipality and year: (i) the \textit{lit area} ($L_{it}$), the number of lit cells; (ii) the \textit{total radiance from the lit area} ($S_{it}$), the radiance summed over those cells; (iii) the \textit{radiance per lit cell} ($I_{it}$), the ratio of the two. Since total radiance is lit area times radiance per lit cell, in logs the effect on total light splits into the effect on how much territory is lit and the effect on how brightly it shines, which identifies the margin that moves. Appendix~\ref{subsubsection:appendix_construction} gives the formal definitions and the construction, while Appendix~\ref{subsubsection:appendix_limits} discusses what the measure captures (Figure~\ref{fig:insets}).

A low threshold is exposed to a drifting background. From 2017, after a change in the processing of the day-night band, cells without artificial light record positive radiance \citep{coesfeld2020reducing}; furthermore, \citet{falchetta2019high} raise their noise floor from 0.25 to 0.35 nW/cm$^2$/sr from that year. At 0.25 nW the share of Italian municipalities with every cell lit, whose lit area therefore cannot grow, moves from 61.2 percent in 2015 to 98.8 percent in 2017 (Table~\ref{tab:sat}). The shift also adds radiance to every cell counted, and since baseline geography and demography predict how many near-threshold cells a municipality holds, covariate adjustment would absorb this artefact as if it were a confounder. We therefore set $\tau = 1$ nW/cm$^2$/sr on the unmasked \texttt{average} band. The producers report a background overwhelmingly below this value \citep{elvidge2017viirs,elvidge2021annual} and place the boundary between lit cells and background near it \citep{elvidge2024statistical}, and on our grid the 95th percentile of radiance in cells with no settlement nearby stays below 1 nW in every year from 2013 to 2022 (Table~\ref{tab:noise_floor}). A higher cut would discard the dim settlement lighting of sparsely populated areas: at 5 nW about one funded municipality in four has no lit cell. Appendix~\ref{subsubsection:appendix_threshold} documents the choice while Section~\ref{section:results_discussion} compares the estimates at 1, 2 and 5 nW.

The product has no complete calendar year before 2013, so the 2011 and 2012 values are extrapolated for each municipality as a linear trend in logs fitted to 2013--2017. Under the estimator of \citet{callaway2021difference} every post-onset comparison is anchored to the year before onset, which is observed (2013 at the earliest), so dropping or refitting the two years leaves the post-onset point estimates of Table~\ref{tab:main} unchanged (Appendix~\ref{subsubsection:appendix_backcast}).

\subsection{Methodology}
\label{subsection:methodology}

We ask whether SNAI 2014--2020 funding left a measurable imprint on the illuminated footprint of funded municipalities, following recent municipal-level evaluations in Italy \citep{fenizia2024organized,di2025regional,lodi2025gender,monturano2025short,mariani2026place}. The outcomes $Y_{i,t}$ are the logs of lit area $L_{it}$, total radiance $S_{it}$ and radiance per lit cell $I_{it}$ (Section~\ref{subsection:measurement}). Treatment is absorbing: $D_{i,t}=1$ from the year $G_i$ in which the first SNAI-funded project in municipality $i$ starts through the last year, $T$. With staggered adoption, a two-way fixed effects regression on $D_{i,t}$ uses already-treated municipalities as controls and is biased when effects vary across cohorts or over time \citep{goodmanbacon2021difference}. We therefore use the estimator of \citet{callaway2021difference}, built on the group-time average treatment effect on the treated,
\begin{equation}
ATT_{(g,t)} = \mathbb{E}[Y_{t}(g) - Y_{t}(0)\mid X, G = g],
\label{eq.att}
\end{equation}
the effect in year $t$ for the cohort first treated in $g$, conditional on baseline covariates $X$. For $t \geq g$ it is identified by comparing the change since $g-1$ with that of municipalities not yet treated at $t$, never-treated included:
\begin{equation}
ATT_{(g,t)} = \mathbb{E}[Y_{t}- Y_{g-1}\mid X, G = g] - \mathbb{E}[Y_{t} - Y_{g-1}\mid X, D_{t}=0, G \neq g].
\label{eq.att_2}
\end{equation}
Group-time effects are aggregated by exposure length $e$,
\begin{equation}
\hat{\theta}_{D}(e) = \sum_{g=2}^{T} 1\{g + e \leq T\} ATT(g, g+e) P(G = g \mid g + e \leq T),
\label{eq.staggered_did_agg}
\end{equation}
and the post-onset ATT of Table~\ref{tab:main} is the simple average of these event-time effects,
\begin{equation}
\hat{\theta}^{O} = \frac{1}{E+1} \sum_{e=0}^{E} \hat{\theta}_{D}(e),
\label{eq.att_post}
\end{equation}
where $E=8$ is the longest horizon observed in the window. We estimate two specifications, without covariates and with the twelve baseline covariates of Section~\ref{subsection:data} in the doubly-robust adjustment. Appendix~\ref{subsection:appendix_estimation} reports the two-way specification, the comparison with never-treated units only and the treatment of covariates in the base period.

The comparison group is restricted to IA municipalities (intermediate, peripheral or ultra-peripheral) of the 2014 classification \citep{dps2014aree,istat_aree_interne}, pilot-area members unfunded by 2021 included (Section~\ref{subsection:data}), which holds eligibility fixed and keeps remoteness comparable.

Identification requires no anticipation and parallel trends. Before the first funded project starts, the lit environment can hardly change, nor are workers and the mostly small firms likely to adjust \citep{mariani2026place}. Administrative capacity to attract and spend funds \citep{arbolino2019factors}, which could allow anticipation, is proxied by three institutional-quality covariates, and with one year of anticipation the post-onset lit-area estimates of Table~\ref{tab:main} are $+0.042$ and $+0.043$.

Parallel trends could fail if pilot-area municipalities differ from other inner municipalities in the institutional quality that helped them form an area, or in remoteness and depopulation, although 99 of the 941 treated municipalities lie outside the 72 pilot areas. Covariates address observable differences, Figure~\ref{fig:es} reports the pre-treatment coefficients, and Table~\ref{tab:designs} re-estimates the effect within the pilot areas.

\subsection{Data}
\label{subsection:data}

The sample comprises the IA municipalities of the 2014--2020 classification \citep{dps2014aree}: 4,055 in the list of \citet{istat_aree_interne}, the only complete one, whereas \citet{nuvap2020strategia} reports the 4,261 given in Section~\ref{section:institutional_setting}. Two of these 4,055, \textit{Montecopiolo} and \textit{Sassofeltrio}, moved from the province of \textit{Pesaro-Urbino} to \textit{Rimini} in 2021 and changed municipal code, so they cannot be matched to the boundary vintage on which the light panel is built. The estimation sample therefore comprises 4,053 municipalities, observed over 2011--2022 ($T=12$).

Treatment comes from the Cohesion-policy monitoring system \href{https://opencoesione.gov.it/it/dati/}{OpenCoesione} \citep{pcm_opencoesione}, updated to October 31, 2025. A municipality is funded if it records at least 1~{\euro} of SNAI 2014--2020 funding, regardless of project-area membership, and its cohort $G_i$ is the year in which its first SNAI project starts. A substantial share of resources was disbursed after 2020. The 941 municipalities with $G_i$ from 2014 to 2021 are treated. The 3,112 controls comprise 3,020 municipalities never funded and 92 whose first project starts in 2022--2024, which are kept as controls in every year of the window. Table~\ref{tab:treated} reports the cumulative number of funded municipalities and their funding by year, and its notes detail how projects recorded above the municipal level are attributed; Figure~\ref{fig:map_treatment} maps treated municipalities by cohort. Section~\ref{section:results_discussion} also redefines treatment by thematic area (Table~\ref{tab:snai_projects_theme}).

The vector $X$ groups geographic, demographic and institutional covariates: surface area, altitude and coastline length (\texttt{coa}) \citep{istat_amdc,istat_geografia}; resident population, the shares of female, foreign and over-65 residents, and the birth, mortality and net migration rates \citep{istat_amdc,istat_demo}; and the three components of the Municipal Administrative Quality Index (MAQI), namely bureaucratic capacity, quality of local politicians, and fiscal and economic performance \citep{cerqua2025municipal}. Covariates are measured in 2013, before the first cohort (for three municipalities, in the first year available), and held fixed in every year, so that they cannot respond to treatment. Income-related variables such as taxable income are excluded, since SNAI interventions can affect them and they would act as bad controls \citep{angrist2009mostly}. Table~\ref{tab:desc} reports descriptive statistics for all variables.\footnote{Notice that coastline length is not among the twelve covariates entered in estimation. See notes to Table~\ref{tab:main}.}

\section{Results}
\label{section:results_discussion} 

This section proceeds in four steps: we first (i) establish the main effect and compare it across radiance thresholds, then (ii) assess its robustness to the choice of comparison group, and finally turn to heterogeneity by (iii) thematic area and by (iv) funding scale.

\begin{figure}[htbp]
	\centering
	\caption{\textit{Event study on threshold-corrected outcomes ($\geq 1$ nW/cm$^2$/sr), horizons up to $e=+6$ (2011--2022)}}
	\includegraphics[width=0.95\linewidth]{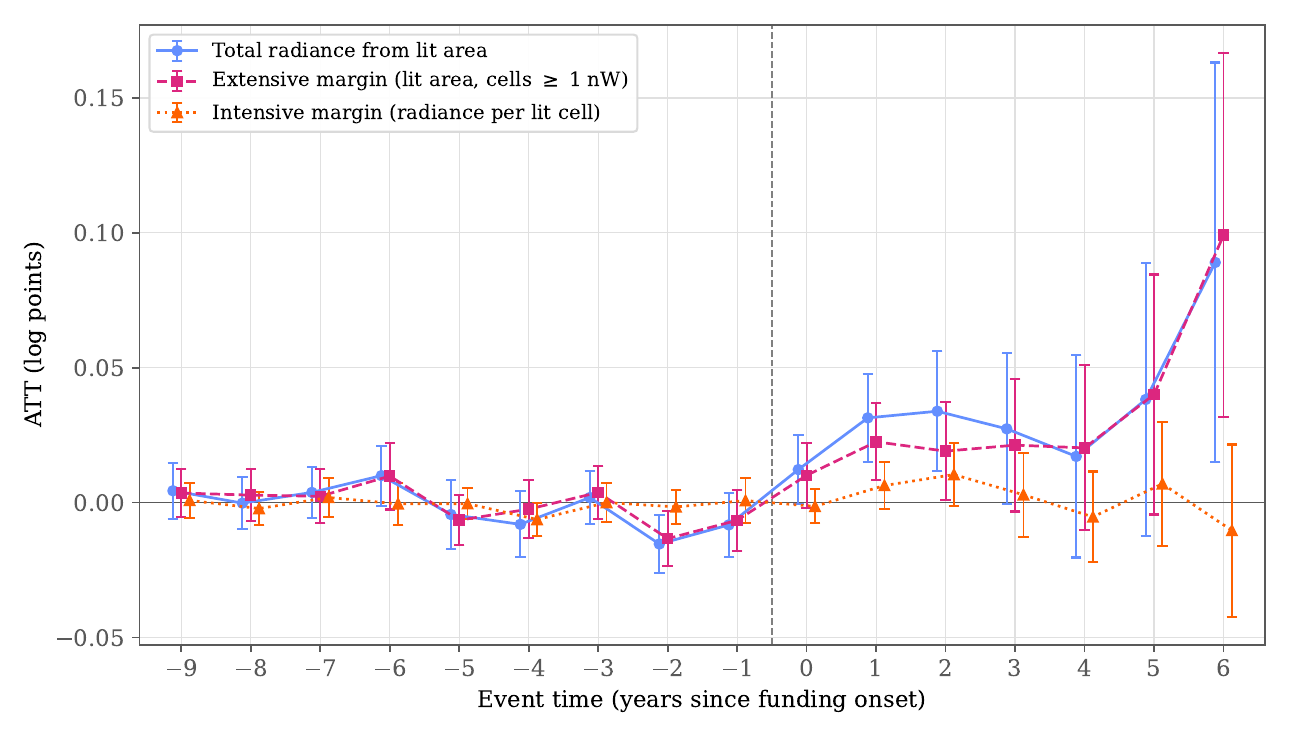}
	\label{fig:es}
	\begin{minipage}{0.92\linewidth}
		\justifying
		\tiny
		\noindent
		Notes: \citet{callaway2021difference} doubly-robust estimates entering the twelve baseline (2013) covariates of Table~\ref{tab:main}, cohorts 2014--2021, comparison group never- and not-yet-treated eligible inner municipalities, 941 treated and 3,112 control municipalities (923 and 3,070 for radiance per lit cell). The horizontal axis stops at $e=+6$; averages over all post-onset horizons are in Table~\ref{tab:main}, and the lit-area series is repeated on the four annual summaries of the composite in Figure~\ref{fig:es_bands_lnlit1}. Bars are pointwise 95\% confidence intervals from a multiplier bootstrap with 300 replications, clustered at the municipality. Across the nine pre-treatment years every coefficient lies within $0.016$ log points of zero; one coefficient per outcome is individually significant at the 5 percent level, at $e=-2$ for total radiance and lit area and at $e=-4$ for radiance per lit cell, and all three are negative.
	\end{minipage}
\end{figure}

Figure~\ref{fig:es} shows the event study for the three threshold-corrected outcomes with the twelve baseline covariates, from $e=-9$ to $e=+6$; averages over all post-onset horizons, including $e=+7$ and $e=+8$, are in Table~\ref{tab:main}. With covariates, the nine pre-treatment coefficients lie within $0.016$ log points of zero, and without them within $0.02$. One pre-treatment coefficient per outcome is individually significant at the 5 percent level, at $e=-2$ for total radiance and lit area and at $e=-4$ for radiance per lit cell, and all three are negative; without covariates, two coefficients on the lit area, three on total radiance and one on radiance per lit cell are individually significant, all negative as well. For total radiance and lit area the coefficients at $e=-2$ and $e=-1$ are both negative, so the pre-treatment path shows a small dip before onset rather than a build-up towards the post-treatment values. Every horizon before onset includes at least one cohort whose comparison involves the reconstructed 2011 or 2012, so the pre-treatment path bears on the extrapolation as well as on the territory (Appendix~\ref{subsubsection:appendix_backcast}). From the onset year, the lit area moves from $+0.010$ at $e=0$ to about $+0.02$ between $e=+1$ and $e=+4$, to $+0.040$ at $e=+5$ and to $+0.099$ at $e=+6$; total radiance follows a similar path, from $+0.012$ to $+0.089$, while radiance per lit cell remains close to zero throughout. The largest coefficients sit at the longest horizons, which rest on the earliest cohorts only, since $e=+6$ is identified by the 2014, 2015 and 2016 cohorts, 173 municipalities.


Table~\ref{tab:main} reports the post-onset averages. The lit area expands by $+0.048$ log points both without covariates and with the twelve baseline covariates, statistically significant at the 5 percent level in both columns. Total radiance from the lit area rises by $+0.037$ and $+0.041$, neither significant at the 5 percent level, and radiance per lit cell moves by $-0.017$ and $-0.008$, indistinguishable from zero in both. The three estimates are approximately additive, since log total radiance is the sum of log lit area and log radiance per cell; they do not add up exactly, $+0.048-0.017=+0.031$ against $+0.037$ without covariates, because total radiance enters as the log of one plus the summed radiance and radiance per lit cell is defined on a smaller panel, and total radiance carries the wider standard error in both columns, $0.024$ against $0.018$ and $0.020$ for the lit area, consistent with the year-to-year volatility of the bright central cells that dominate it. The response is concentrated at the dim margin, which weighs little in total radiance and fully in the lit area, so that the lit area is the informative margin. Its point estimate is the same under the two conditioning sets and does not depend on how the annual composite summarises the year. Rebuilt on the median, the maximum and the minimum of the monthly radiances that the producer distributes alongside their average, the lit-area estimate is positive on all four summaries, significant at the 5 percent level over all post-onset horizons on the average, the median and the maximum, and over horizons 0 to +6 on all four; the median reproduces the baseline, while the maximum, whose pre-treatment path is less flat, gives an upper envelope (Appendix~\ref{subsection:bands}).

\begin{table}[htbp]
\centering
\caption{\textit{Post-onset ATT on threshold-corrected outcomes (2011--2022)}}
\label{tab:main}
\small
\begin{tabular}{lcc}
\toprule
 & \textbf{(1)} & \textbf{(2)} \\
 & \textit{No covariates} & \textit{Baseline covariates} \\
\midrule
\multicolumn{3}{l}{\textit{Post-onset ATT (log points)}} \\[2pt]
Total radiance from lit area & $+0.037$ & $+0.041^{*}$ \\
 & $(0.024)$ & $(0.024)$ \\[2pt]
Lit area (cells $\geq 1$ nW/cm$^2$/sr) & $+0.048^{***}$ & $+0.048^{**}$ \\
 & $(0.018)$ & $(0.020)$ \\[2pt]
Radiance per lit cell & $-0.017$ & $-0.008$ \\
 & $(0.013)$ & $(0.013)$ \\[2pt]
\midrule
Treated municipalities & 941 & 941\\
Control municipalities & 3,112 & 3,112\\
\bottomrule
\end{tabular}
\begin{minipage}{0.85\linewidth}~\\
\justifying
\tiny
\noindent
Notes: \citet{callaway2021difference} doubly-robust estimates, cohorts 2014--2021, comparison group never- and not-yet-treated eligible inner municipalities, no anticipation. Bootstrap standard errors with 300 replications, clustered at the municipality, in parentheses. Outcomes are measured above a stable radiance threshold of 1 nW/cm$^2$/sr. Column (2) enters twelve baseline (2013) covariates in the doubly-robust adjustment, those listed in Section~\ref{subsection:data} except the coastal variable (\texttt{coa}), which is zero for every municipality of the 2014, 2015 and 2018 cohorts, so that entering it would make the propensity score singular in those cohorts and drop them from the adjusted estimates. Each estimate is the simple average of the event-time effects over all post-onset horizons observed in the window, $e=0,\dots,+8$. Radiance per lit cell is defined only for municipalities with at least one lit cell in every year, so that for this outcome the panel retains 923 treated and 3,070 control municipalities. Standard errors are computed with a fixed random seed and coincide with those of the rows for the average summary in Panel A of Table~\ref{tab:bands}.  $^{***}$ $p<0.01$, $^{**}$ $p<0.05$, $^{*}$ $p<0.10$. \\
\end{minipage}
\end{table}

Table~\ref{tab:thresholds} and Figure~\ref{fig:es_thresholds} repeat the estimation counting a cell as lit above 2 and 5 nW/cm$^2$/sr instead of 1 nW. With the twelve baseline covariates the expansion of the lit area is of similar size at the three thresholds, $+0.048$, $+0.051$ and $+0.053$ log points over all post-onset horizons, all significant at the 5 percent level. Without covariates, it is about half as large at the two higher thresholds, $+0.023$ and $+0.024$, and not significant, so the agreement between the two specifications observed at 1 nW does not extend to them. Over horizons $0$ to $+6$ the lit-area estimates are positive and significant at the 5 percent level at all three thresholds and in both specifications, between $+0.023$ and $+0.051$. The thresholds do not measure the same cells: at 5 nW about one funded municipality in four has no lit cell at all, against about one in a hundred at 1 nW. Radiance per lit cell stays close to zero over all horizons at every threshold; over horizons $0$ to $+6$ it is small and positive at 2 and 5 nW, between $+0.013$ and $+0.016$ log points and significant at the 5 percent level in three of the four estimates, where it is defined for fewer municipalities, 874 and 612 of the 941 treated. We keep 1 nW as the main threshold for the reasons given in Section~\ref{subsection:measurement}: it lies above the background level of the product while retaining the dim settlement lighting on which the extensive margin depends.

The concern raised by Table~\ref{tab:main} is that funded municipalities are selected on decline, so we re-estimate the effect against three comparison groups that trade comparability against statistical power, reported in Table~\ref{tab:designs}. From narrowest to widest: the 143 unfunded members of the pilot areas, which hold the selection procedure fixed; the 3,112 unfunded IA municipalities of the baseline, which hold eligibility fixed; and the 6,787 Italian municipalities of every SNAI class, which hold nothing fixed and let the covariates do all the work.

The tightest comparison holds the selection fixed. Among the 1,060 municipalities of the 72 pilot areas, 985 belong to the inner belts, 842 of which received funding, while 143 did not, either because funds never arrived (79) or because they arrived from 2022 onward (64). Both groups passed the same selection procedure, so the comparison isolates the arrival of money from the process that designated the areas. The lit area expands by $+0.078$ log points without covariates and $+0.079$ with them, roughly 1.6 times the baseline estimate, with radiance per lit cell again indistinguishable from zero. The cost is precision: 143 controls widen the standard errors, and the composition of that group deserves caution, since municipalities whose funds never arrived may differ in administrative capacity. The baseline broadens the comparison group to the full eligible universe, which multiplies the controls twentyfold and yields the policy-relevant estimand at $+0.048$. The third design abandons eligibility altogether and compares the same treated municipalities with all Italian municipalities, hubs and belt included. Without covariates, the estimate inflates to $+0.097$ and radiance per lit cell turns sharply negative, exactly as the measurement discussion of Section~\ref{subsection:measurement} anticipates: urban controls have lit areas that are already saturated, so their extensive margin cannot grow, and their intensity is rising. Once the baseline covariates absorb that compositional imbalance, the estimate returns to the range of the eligible-sample design, at $+0.054$.

The lit-area effect is positive and statistically significant in all six rows, across comparison groups ranging from 143 to 6,787 units. And the tightest comparison runs against the selection objection: holding the selection procedure fixed yields a larger estimate than the baseline, $+0.078$ and $+0.079$ against $+0.048$, not a smaller one.

\begin{table}[H]
\centering
\caption{\textit{Post-onset ATT under alternative comparison groups (2011--2022)}}
\label{tab:designs}
\small
\resizebox{\textwidth}{!}{%
\begin{tabular}{lccccc}
\toprule
\textbf{Specification} & \makecell{\textbf{N}\\\textbf{treated}} & \makecell{\textbf{N}\\\textbf{controls}} & \makecell{\textbf{Total radiance}\\\textbf{from lit area}} & \makecell{\textbf{Lit area}\\\textbf{($\geq 1$ nW)}} & \makecell{\textbf{Radiance per}\\\textbf{lit cell}} \\
\midrule
\multicolumn{6}{l}{\textbf{Within-strategy}} \\[1pt]
\quad No covariates & 842 & 143 & $+0.095^{***}$ $(0.034)$ & $+0.078^{***}$ $(0.029)$ & $+0.010$ $(0.017)$ \\
\quad 12 covariates & 842 & 143 & $+0.077$ $(0.048)$ & $+0.079^{**}$ $(0.034)$ & $-0.012$ $(0.023)$ \\
\addlinespace[7pt]
\multicolumn{6}{l}{\textbf{Baseline: funded inner municipalities}} \\
\quad No covariates & 941 & 3,112 & $+0.037$ $(0.024)$ & $+0.048^{***}$ $(0.018)$ & $-0.017$ $(0.013)$ \\
\quad 12 covariates & 941 & 3,112 & $+0.041^{*}$ $(0.024)$ & $+0.048^{**}$ $(0.020)$ & $-0.008$ $(0.013)$ \\
\addlinespace[7pt]
\multicolumn{6}{l}{\textbf{Full-Italy control pool}} \\
\quad No covariates & 941 & 6,787 & $+0.045^{*}$ $(0.026)$ & $+0.097^{***}$ $(0.020)$ & $-0.058^{***}$ $(0.011)$ \\
\quad 12 covariates & 941 & 6,787 & $+0.034$ $(0.024)$ & $+0.054^{***}$ $(0.019)$ & $-0.020$ $(0.013)$ \\
\bottomrule
\end{tabular}%
}
\begin{minipage}{0.98\linewidth}~\\
\justifying
\tiny
\noindent
Notes: each block re-estimates the design of Table~\ref{tab:main}, changing only the definition of the treated group and of the comparison group, as stated in the block description. Estimates average the event-time effects over all post-onset horizons, as in columns (1) and (2) of Table~\ref{tab:main}, and the baseline block reproduces those two columns. Outcomes are measured above the 1 nW/cm$^2$/sr threshold; window 2011--2022. Bootstrap standard errors with 300 replications, clustered at the municipality, in parentheses. $^{***}$ $p<0.01$, $^{**}$ $p<0.05$, $^{*}$ $p<0.10$. \\
\end{minipage}
\end{table}

These results add a spatial dimension to the emerging evidence on the effects of the SNAI. More broadly, \cite{bartik2020using} argues that labour-market failures provide an economic rationale for place-based policies, whose returns may be greater in economically disadvantaged areas than in more prosperous ones. In the SNAI context, \cite{monturano2025short} finds that initial funding inflows increased the number of business establishments in treated municipalities and generated positive spillovers in neighbouring areas, while \cite{di2025regional} documents an increase in tourist nights, primarily driven by domestic tourism and concentrated in specific territorial contexts.

The expansion of the illuminated footprint is consistent with these findings and suggests a broader, potentially earlier manifestation of local economic activation. The observed pattern points to changes at the margins of existing settlements, without a comparable increase in light intensity within their established cores. Nighttime light indicators also provide a complementary perspective on outcomes that conventional labour-market measures may overlook. For example, although \cite{mariani2026place} finds no detectable employment effects, where data availability restricts the analysis to employee labuor market, potential effects on self-employment may remain unobserved. Our indicators may capture spatial changes associated with such activities, although they cannot identify their contribution separately or directly measure employment creation.

The absence of an intensive-margin response is also consistent with evidence of limited effects on population, employment and accessibility \citep{monturano2025short,bergantino2026far,mariani2026place}. Taken together, these findings are compatible with an adjustment process in which SNAI funding initially generates localised changes in business activity, tourism and the spatial distribution of activity. Such changes may precede broader demographic and structural transformations, but do not necessarily translate into them. This interpretation suggests that early spatial responses can coexist with limited aggregate effects, while leaving open whether more substantial transformations will emerge over time.

\subsection{Heterogeneity of the policy}
\label{subsection:heterogeneity}

To explore whether the aggregate effect masks relevant heterogeneity, we extend the baseline analysis along two dimensions. First, we assess whether the estimated impact is associated with specific thematic areas of intervention. Second, we examine whether the response varies with the amount of funding received, considering both total and per-capita allocations under different thresholds.

Table \ref{tab:sector} investigates heterogeneity across the eleven thematic areas. For each area, we re-estimate the design twice: once defining treatment as the arrival of funding in that area alone, and once defining it as the arrival of any funding except that area.

Excluding any single thematic area leaves the lit-area estimate between $+0.029$ and $+0.059$ log points, statistically significant in ten of the eleven cases and, in those ten, indistinguishable from the baseline. The single exception is digital networks and services, at $+0.029$ with a standard error of $0.024$, and it is also the one area whose own-area estimate on the lit area is positive and significant, so that removing it appears to subtract part of the response rather than leave it intact. No thematic area carries the aggregate response on its own, and the effect follows the arrival of money rather than its composition. The only visible movement in the other direction is that excluding transport strengthens the estimate slightly, which is consistent with the weak own-effect of that area.

The first set of rows, defining treatment by a single thematic area, is more dispersed. Its comparison group is contaminated, since municipalities without funding in a given area may well receive funding in others, and municipalities funded across several areas enter multiple rows. With that caveat, digital networks and services stand out with an expansion of the lit area of $+0.047$ accompanied by a decline in radiance per cell of $-0.028$, a pattern consistent with lighting and network works that extend coverage at lower brightness per cell.

The absence of a single thematic area driving the aggregate response also qualifies the mechanisms identified in previous studies. \cite{di2025regional}, focusing specifically on tourism outcomes, finds that the increase in tourist nights is associated mainly with short-term promotional expenditure rather than with structural projects. Our results indicate that the broader expansion of the illuminated footprint cannot be attributed to one expenditure category alone and may reflect the combined territorial effects of different interventions. This interpretation is consistent with the integrated design of the SNAI, in which investments in services, infrastructure and local development are intended to operate jointly, and with the broader shift from sectoral interventions towards integrated development projects that combine hard and soft capital in modern place-based policy \citep{garcilazo2025case}. 

\begin{table}[H]
\centering
\caption{\textit{Post-onset ATT by thematic area of spending (2011--2022)}}
\label{tab:sector}
\footnotesize
\resizebox{\textwidth}{!}{%
\begin{tabular}{llcccc}
\toprule
\textbf{Thematic area} & \textbf{Treatment} & \textbf{N treated} & \makecell{\textbf{Total radiance}\\\textbf{from lit area}} & \makecell{\textbf{Lit area}\\\textbf{($\geq 1$ nW)}} & \makecell{\textbf{Radiance per}\\\textbf{lit cell}} \\
\midrule
\multicolumn{2}{l}{\textit{Any funding (baseline)}} & 941 & $+0.037$ $(0.024)$ & $+0.048^{**}$ $(0.021)$ & $-0.017$ $(0.013)$ \\ \midrule
Administrative Capacity & only & 329 & $+0.029$ $(0.029)$ & $+0.026$ $(0.027)$ & $+0.001$ $(0.008)$ \\
 & excluding & 915 & $+0.034$ $(0.026)$ & $+0.046^{**}$ $(0.018)$ & $-0.017$ $(0.012)$ \\[2pt]
Business Competitiveness & only & 259 & $+0.014$ $(0.023)$ & $-0.017$ $(0.023)$ & $+0.035^{***}$ $(0.010)$ \\
 & excluding & 901 & $+0.035$ $(0.026)$ & $+0.049^{**}$ $(0.021)$ & $-0.020^{*}$ $(0.012)$ \\[2pt]
Culture and Tourism & only & 301 & $-0.053^{*}$ $(0.030)$ & $-0.020$ $(0.026)$ & $-0.033$ $(0.025)$ \\
 & excluding & 931 & $+0.035$ $(0.026)$ & $+0.048^{**}$ $(0.021)$ & $-0.019$ $(0.013)$ \\[2pt]
Digital Networks and Services & only & 246 & $+0.018$ $(0.018)$ & $+0.047^{**}$ $(0.019)$ & $-0.028^{***}$ $(0.009)$ \\
 & excluding & 893 & $+0.023$ $(0.029)$ & $+0.029$ $(0.024)$ & $-0.012$ $(0.014)$ \\[2pt]
Education and Training & only & 294 & $+0.173$ $(0.130)$ & $+0.107$ $(0.109)$ & $+0.055^{**}$ $(0.023)$ \\
 & excluding & 935 & $+0.036$ $(0.026)$ & $+0.049^{**}$ $(0.023)$ & $-0.018$ $(0.011)$ \\[2pt]
Employment and Work & only & 146 & $+0.268^{*}$ $(0.137)$ & $+0.163$ $(0.105)$ & $+0.076^{***}$ $(0.025)$ \\
 & excluding & 938 & $+0.031$ $(0.027)$ & $+0.045^{**}$ $(0.020)$ & $-0.018$ $(0.013)$ \\[2pt]
Energy & only & 93 & $-0.051$ $(0.039)$ & $+0.022$ $(0.024)$ & $-0.072^{*}$ $(0.038)$ \\
 & excluding & 935 & $+0.037$ $(0.026)$ & $+0.049^{**}$ $(0.020)$ & $-0.018$ $(0.014)$ \\[2pt]
Environment & only & 214 & $-0.031$ $(0.021)$ & $-0.026$ $(0.024)$ & $-0.004$ $(0.016)$ \\
 & excluding & 937 & $+0.038$ $(0.027)$ & $+0.051^{**}$ $(0.020)$ & $-0.019$ $(0.012)$ \\[2pt]
Research and Innovation & only & 40 & $+0.065^{**}$ $(0.029)$ & $+0.030$ $(0.024)$ & $+0.038^{**}$ $(0.018)$ \\
 & excluding & 940 & $+0.037$ $(0.024)$ & $+0.048^{**}$ $(0.020)$ & $-0.017$ $(0.013)$ \\[2pt]
Social Inclusion and Health & only & 321 & $+0.147$ $(0.099)$ & $+0.126$ $(0.120)$ & $+0.035$ $(0.021)$ \\
 & excluding & 899 & $+0.038^{*}$ $(0.022)$ & $+0.048^{**}$ $(0.020)$ & $-0.017$ $(0.012)$ \\[2pt]
Transport and Mobility & only & 473 & $-0.037$ $(0.029)$ & $+0.012$ $(0.019)$ & $-0.050^{***}$ $(0.019)$ \\
 & excluding & 860 & $+0.054^{*}$ $(0.032)$ & $+0.059^{**}$ $(0.025)$ & $-0.011$ $(0.014)$ \\
\bottomrule
\end{tabular}%
}
\begin{minipage}{0.98\linewidth}~\\
\justifying
\tiny
\noindent
Notes: each row re-estimates the design of Table~\ref{tab:main}, without covariates, changing only the definition of treatment. ``Only'' defines treatment as the first arrival of funding in the stated thematic area; ``excluding'' as the first arrival of any SNAI funding other than that area. The eleven thematic areas partition total funding exactly (Table~\ref{tab:snai_projects_theme}). Own-area estimates are descriptive, since their comparison group includes municipalities funded in other areas and multi-area municipalities enter several rows. The baseline row is re-estimated within the same bootstrap run as the thematic rows. In that run, unlike in Table~\ref{tab:main}, municipalities whose first arrival of funding, as defined in each row, falls in 2022 enter as a cohort treated in 2022 rather than as controls, and are left out of the count of treated municipalities; in the baseline row, they are the 51 municipalities first funded in 2022. The point estimates of the baseline row coincide with those of column (1) of Table~\ref{tab:main} to the third decimal, while its standard error on the lit area, $0.021$ against $0.018$, makes the lit-area estimate significant at the 5 rather than the 1 percent level. Bootstrap standard errors with 300 replications, clustered at the municipality, in parentheses. $^{***}$ $p<0.01$, $^{**}$ $p<0.05$, $^{*}$ $p<0.10$. \\
\end{minipage}
\end{table}

\begin{table}[H]
\centering
\caption{\textit{Post-onset ATT by quartile of funding received (2011--2022)}}
\label{tab:dose}
\setlength{\tabcolsep}{4pt}
\renewcommand{\arraystretch}{1.15}
\scriptsize
\begin{tabular}{@{}l l ccc@{}}
\toprule
 & & \makecell{\textbf{Total radiance}\\\textbf{from lit area}}
   & \makecell{\textbf{Lit area}\\\textbf{($\geq 1$ nW)}}
   & \makecell{\textbf{Radiance per}\\\textbf{lit cell}} \\
\midrule
\multicolumn{5}{@{}l}{\textit{Panel A. Quartiles of cumulative funding (end-2022)}} \\[2pt]
\makecell[l]{Q1\\[-1pt]\tiny(N=236; 13\,k\euro)}    & no cov.  & $-0.023\;(0.045)$ & $-0.052^{*}\;(0.029)$ & $+0.016\;(0.016)$ \\
                                                     & 13 cov.  & $-0.006\;(0.034)$ & $+0.012\;(0.024)$ & $-0.011\;(0.011)$ \\[3pt]
\makecell[l]{Q2\\[-1pt]\tiny(N=235; 99\,k\euro)}    & no cov.  & $+0.199^{*}\;(0.111)$ & $+0.145^{*}\;(0.081)$ & $+0.049^{**}\;(0.020)$ \\
                                                     & 13 cov.  & $-0.015\;(0.031)$ & $-0.004\;(0.023)$ & $-0.003\;(0.019)$ \\[3pt]
\makecell[l]{Q3\\[-1pt]\tiny(N=235; 335\,k\euro)}   & no cov.  & $-0.001\;(0.042)$ & $+0.030\;(0.031)$ & $-0.035^{*}\;(0.020)$ \\
                                                     & 13 cov.  & $+0.025\;(0.032)$ & $+0.016\;(0.027)$ & $+0.004\;(0.013)$ \\[3pt]
\makecell[l]{Q4\\[-1pt]\tiny(N=235; 1{,}508\,k\euro)} & no cov. & $-0.013\;(0.020)$ & $+0.028\;(0.018)$ & $-0.040^{**}\;(0.016)$ \\
                                                     & 13 cov.  & $+0.026\;(0.021)$ & $+0.027\;(0.018)$ & $+0.002\;(0.010)$ \\
\midrule
\multicolumn{5}{@{}l}{\textit{Panel B. Quartiles cumulative of per-capita funding}} \\[2pt]
\makecell[l]{Q1\\[-1pt]\tiny(N=236; 20\,\euro/res.)}  & no cov. & $-0.083^{***}\;(0.030)$ & $-0.078^{***}\;(0.024)$ & $-0.010\;(0.014)$ \\
                                                     & 13 cov.  & $-0.033\;(0.027)$ & $-0.015\;(0.023)$ & $-0.007\;(0.015)$ \\[3pt]
\makecell[l]{Q2\\[-1pt]\tiny(N=235; 77\,\euro/res.)}  & no cov. & $+0.095^{***}\;(0.028)$ & $+0.066^{***}\;(0.022)$ & $+0.026^{**}\;(0.011)$ \\
                                                     & 13 cov.  & $+0.060^{*}\;(0.033)$ & $+0.060^{**}\;(0.025)$ & $+0.006\;(0.014)$ \\[3pt]
\makecell[l]{Q3\\[-1pt]\tiny(N=235; 234\,\euro/res.)} & no cov. & $+0.114\;(0.079)$ & $+0.113^{*}\;(0.064)$ & $-0.009\;(0.022)$ \\
                                                     & 13 cov.  & $+0.038^{**}\;(0.019)$ & $+0.061^{***}\;(0.021)$ & $-0.026^{**}\;(0.012)$ \\[3pt]
\makecell[l]{Q4\\[-1pt]\tiny(N=235; 847\,\euro/res.)} & no cov. & $+0.001\;(0.026)$ & $+0.022\;(0.022)$ & $-0.023\;(0.019)$ \\
                                                     & 13 cov.  & $+0.066^{**}\;(0.034)$ & $+0.030\;(0.022)$ & $+0.044^{**}\;(0.019)$ \\
\bottomrule
\end{tabular}

\vspace{4pt}
\begin{minipage}{0.78\linewidth}
\justifying\tiny\noindent
Notes: each row re-estimates the design of Table~\ref{tab:main}, without and with the thirteen baseline covariates, restricting the treated group to the stated quartile of the funding distribution among treated municipalities; the comparison group is unchanged and treated municipalities in other quartiles are excluded. Per-capita funding divides cumulative funding by 2013 resident population. Quartiles group different municipalities by construction, so the comparison across quartiles is descriptive. Bootstrap standard errors with 300 replications, clustered at the municipality, in parentheses. $^{***}$ $p<0.01$, $^{**}$ $p<0.05$, $^{*}$ $p<0.10$.\\
\end{minipage}
\end{table}

Table \ref{tab:dose} shows no systematic heterogeneity by total funding, whereas per-capita funding reveals a nonlinear pattern. After covariate adjustment, the lowest quartile displays no detectable effect on lit area, while the second and third quartiles show increases of approximately 6 percent. The estimate becomes smaller and statistically insignificant in the highest quartile. These results suggest that the territorial response requires sufficient funding intensity but does not increase proportionally beyond that level. The nonlinear relationship with funding intensity provides a further connection with the previous evidence. \cite{monturano2025short} stress that the early implementation of the SNAI was constrained by the limited share of programmed resources that had actually reached municipalities, while \cite{mariani2026place} shows that policy effects depend strongly on local conditions, including administrative capacity and productive structure. Our findings complement these results by identifying funding scale as an additional condition for effectiveness. \cite{andini2016local} find a similar pattern for Italy’s \textit{Contratti di Programma}, whose small effects appear larger at intermediate funding intensities, while the growth returns to Structural Funds stop rising beyond a certain intensity and depend on human capital and institutional quality (Section~\ref{subsection:place_based}).
In IAs, very small allocations may be insufficient to cover the fixed costs and coordination requirements of activating new economic activity, while beyond a certain scale the limited administrative and productive capacity of small, sparsely populated municipalities may prevent additional resources from translating into a proportional expansion of the illuminated area.

\begin{figure}[H]
	\centering
	\caption{\textit{Event studies by group of per-capita funding}}
	\includegraphics[width=0.98\linewidth]{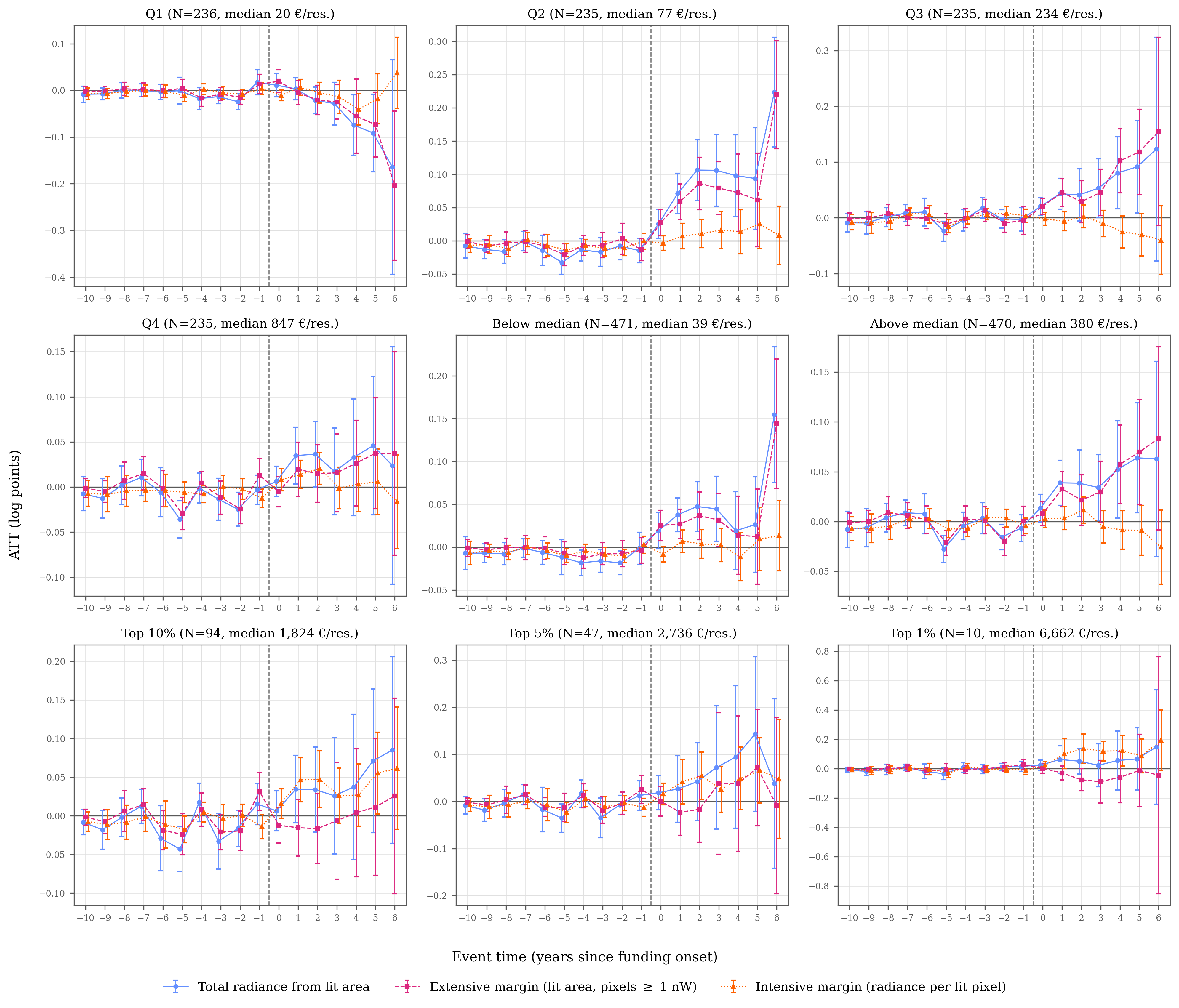}
	\label{fig:qpc}
	\begin{minipage}{0.92\linewidth}
		\justifying
		\tiny
		\noindent
		Notes: each panel reports an event study estimated as in Figure~\ref{fig:es} but without covariates, restricting the treated group to the stated group of the per-capita funding distribution and excluding the other municipalities first funded in 2014--2021. Unlike in Figure~\ref{fig:es}, the 51 municipalities first funded in 2022 enter as a cohort treated in 2022: they serve as not-yet-treated controls up to 2021 and contribute to the coefficients from $e=-10$ to $e=0$ in every panel, with the coefficient at $e=-10$ resting on them alone. Sample medians are 124 euro per resident, and the 90th, 95th and 99th percentiles are 1,042, 1,813 and 4,798 euro per resident respectively. Bars are 95\% bootstrap confidence intervals clustered at the municipality. The top 1\% group contains ten municipalities and is reported for completeness. \\
	\end{minipage}
\end{figure}


\section{Conclusions}
\label{section:conclusions}

This paper examined whether the Italian National Strategy for Inner Areas produced a measurable change in the spatial distribution of local economic activity. To do so, we combined a newly assembled municipal-level dataset on SNAI funding with annual VIIRS nighttime-light data for the period 2011–2022, in which the two opening years are reconstructed from the observed 2013--2017 path. We distinguished between the expansion of illuminated territory and changes in the brightness of already illuminated areas, and estimated the effects of the programme using a staggered difference-in-differences design comparing 941 funded municipalities with eligible IA municipalities that had not yet received, or never received, SNAI funding.

The results show that SNAI funding increased the illuminated area of treated municipalities by approximately 5 percent. The effect emerges after the arrival of funding and grows progressively over the following six years. By contrast, we find no corresponding increase in radiance per lit cell. The different programs therefore appears to have affected the extensive rather than the intensive margin: newly illuminated cells emerged mainly around the edges of existing settlements, while their established cores did not become brighter. 

The main result is robust to the inclusion of the baseline covariates and to the choice of comparison group: the expansion of the lit area remains positive and statistically significant whether or not the covariates enter the doubly-robust adjustment, and whether the funded members of the pilot areas are compared with the unfunded ones, or the 941 treated municipalities with the full eligible IA universe or with the Italy-wide pool of 6,787 municipalities. With the baseline covariates it is of similar size at thresholds of 2 and 5 nW/cm$^2$/sr, and it is positive when the annual composite is summarised by the median, the maximum or the minimum of the monthly radiances instead of their average, although on the minimum it is significant only over horizons 0 to +6 (Appendix~\ref{subsection:bands}). It also does not appear to be driven by any single category of expenditure: removing individual thematic areas leaves the estimated expansion of the illuminated footprint positive in every case and statistically significant in ten of the eleven, the exception being digital networks and services. Funding intensity, however, matters. The response becomes detectable only after expenditure reaches a minimum level per resident, while larger allocations beyond the middle of the funding distribution do not produce proportionally greater effects. 

These findings provide evidence that a major place-based programme can activate previously underused portions of marginal territories rather than simply reinforce existing local centres. They also carry a practical implication for policy design: spreading limited resources too thinly may fail to produce a detectable territorial response, whereas reaching a critical investment scale appears more important than concentrating expenditure in a particular thematic area.

Taking into account the methodological contribution of this paper, the analysis highlights the potential of satellite data as a tool for the timely monitoring of place-based policies. Unlike conventional socioeconomic indicators, which are often released with considerable delay and at relatively aggregated spatial scales, nighttime-light observations are available frequently, consistently and at fine geographical resolution. Once combined with georeferenced and regularly updated information on policy implementation, they could support a monitoring system capable of detecting where and when interventions begin to leave a visible territorial footprint, well before official economic and demographic statistics become available. Such evidence cannot replace conventional impact evaluation or welfare indicators, but it can complement them by providing policymakers with an early signal of implementation, helping to identify areas in which resources are producing observable change and those in which corrective action may be needed. In this sense, satellite data could move the evaluation of territorial policies closer to real time and transform it from an exercise conducted mainly after programme completion into a tool that also informs policy during implementation. Two qualifications bound this prospect, and recent remote-sensing work documents both. The first concerns frequency. Annual composites record the year and not the month, and the light signal itself is slow to move: \citet{voncarnap2026satellite} build a measure of periodic-market activity from daytime PlanetScope imagery at a resolution of 3.1 m, available with a median average revisit interval of thirty hours, and track 1,776 marketplaces across Ethiopia outside its capital at up-to-weekly frequency between 2017 and 2024, on the stated ground that night-time light emissions do not reliably register short-term change and respond only slowly to changing economic conditions. For a programming cycle that unfolds over years this limit is not binding, but it marks the distance between the timeliness claimed here and the timeliness now attainable from satellites. The second qualification concerns coverage, because a light-based system observes only what rises above the background of the product: 63 percent of the marketplaces they detect lie in locations with no measurable light emission between 2022 and 2024 in the annual VIIRS composites of \citet{elvidge2021annual}, the product from which our outcomes are also built. The drifting background documented in Section~\ref{subsection:measurement} is therefore not only a source of error over time but also a boundary on where an instrument of this kind can see anything at all, and a fixed threshold of 1 nW/cm$^2$/sr, set above the background of the product, draws that boundary more tightly, since activity that does not lift a cell above it leaves no trace in the lit area.

The interpretation nevertheless requires caution. Nighttime lights capture the physical footprint of human and economic activity, not improvements in income, employment, productivity or residents’ welfare. An expansion of illuminated territory should therefore be understood as evidence that the programme left a tangible spatial imprint, rather than as a comprehensive measure of local development. Linking this footprint to demographic, labour-market and welfare outcomes is the natural next step for assessing whether the territorial changes documented here translated into lasting improvements in the conditions of Italy’s inner areas. A further limit is normative rather than empirical: we ask whether public funding left a visible mark on peripheral territory, not whether sustaining activity in dispersed locations is an efficient spatial configuration, and theory offers no settled presumption on the latter. \citet{anas2004vanishing} shows that the new economic geography framework, once stripped of the constraints imposed by geography, implies that de-urbanisation rather than concentration is the efficient path as population grows, and reads the agglomeration such models normally deliver as an artefact of the spatial constraints placed upon them.

\pagebreak

\bibliographystyle{apalike-etal-in-italics} 
\bibliography{bib}

\pagebreak

\section*{Appendix}
\label{section:appendix} 

\setcounter{section}{0}
\setcounter{subsection}{0}
\setcounter{subsubsection}{0}
\renewcommand{\thesubsection}{A.\arabic{subsection}}
\renewcommand{\thesubsubsection}{A.\arabic{subsection}.\arabic{subsubsection}}
\setcounter{table}{0}
\renewcommand{\thetable}{A\arabic{table}}
\setcounter{figure}{0}
\renewcommand{\thefigure}{A\arabic{figure}}
\setcounter{equation}{0}
\renewcommand{\theequation}{a\arabic{equation}}

\subsection{Night-light data: sources, threshold, and sensor continuity}
\label{subsection:appendix_lights}

\subsubsection{The families of night-light products}
\label{subsubsection:appendix_products}

Four families of night-light data are in circulation, and they are not interchangeable. The distinction matters here because the choice of product determines which margin of the light response can be measured at all.

The first is the \textbf{DMSP-OLS} series. Its images are distributed on a 30 arc-second grid, roughly one kilometre at Italian latitudes, and its values are six-bit digital numbers bounded between 0 and 63 rather than physical radiances. Three consequences follow. The sensor has no on-board calibration, so levels are not comparable across satellites or years without adjustment. Its point-spread function is wide, so light is smeared over neighbouring cells, which \cite{abrahams2018deblurring} address with an explicit de-blurring correction. And the six-bit ceiling truncates the brightest areas, the pathology that motivates the top-coding correction of \cite{bluhm2022top}. Taken together, blurring and truncation compress the measured distribution, which is why \cite{zhang2023remotely} find that DMSP data understate spatial inequality.

The second is the \textbf{VIIRS} day-night band, distributed for annual analysis as the VNL composites \citep{elvidge2017viirs,elvidge2021annual}. Images are on a 15 arc-second grid, about 463 metres north to south and narrower east to west at Italian latitudes, and values are calibrated radiances in nW/cm$^2$/sr. There is no six-bit ceiling. The annual average radiance is computed from monthly composites of observations free of sunlight, moonlight and cloud, and biomass burning, aurora and the background are zeroed out only in the masked bands of the product \citep{elvidge2021annual}. We use the unmasked \texttt{average} band. The masked band is not a substitute, since its background mask discards dim lighting such as dimly lit roads \citep{elvidge2021annual}, which is the margin we measure, and the rules that build the mask change between the two releases of the product that our panel joins \citep{eog2023vnlreadme}.

The third family consists of \textbf{harmonised DMSP-VIIRS} series, of which \cite{li2020harmonized} is the most widely used. These convert VIIRS radiances back onto the DMSP digital-number scale in order to obtain a single long panel from 1992 onward. The gain is length; the cost is that the resulting series is expressed in a bounded, uncalibrated unit and that two physically different sensors are joined inside it.

The fourth consists of \textbf{VIIRS-like reconstructions}, which run the conversion in the opposite direction, using cross-sensor calibration to express the DMSP era in radiance-like units at VIIRS resolution. They inherit the same junction problem in a less visible form, since the output looks like radiance throughout.

\subsubsection{Why the VIIRS annual composites for this design}
\label{subsubsection:appendix_viirs}

Our question is which margin of the light response moves, and answering it requires a product with three properties that only the VIIRS composites possess jointly.

The first is \textbf{physical units}. Separating an expansion of the illuminated footprint from a brightening of existing centres requires a threshold that means the same thing in every year and in every municipality. A threshold can be fixed only if the underlying quantity is a calibrated radiance. On a digital-number scale bounded a cut-off is a position in a compressed and satellite-specific distribution, not a physical brightness.

The second is \textbf{the absence of a ceiling}. The intensive margin is the average radiance of the lit cells. On a truncated product, that average is mechanically bounded above, and any brightening of already bright cells is invisible by construction. Reporting a null intensive-margin effect would then be uninformative, since the product could not have shown one.

The third is \textbf{resolution}. IA municipalities are small and dimly lit. At one kilometre, the median municipality in our sample holds a few tens of cells and the lit-cell count moves in coarse jumps; at 500 metres it holds four times as many, which is what makes a count-based extensive margin usable at this scale. This is also why we do not follow the practice, common at regional scale, of collapsing a unit into a single luminosity sum: at municipal scale the sum is dominated by a handful of bright central cells whose year-to-year volatility is unrelated to the policy.

\subsubsection{The choice of the threshold}
\label{subsubsection:appendix_threshold}

Given a calibrated radiance, a threshold still has to be chosen, and three families of choice are available. We use a \textbf{fixed physical threshold}, the same value in every year and for every municipality. The two alternatives are rejected for reasons that are worth stating, because both appear in applied work.

A \textbf{within-year percentile} of the radiance distribution, such as the top decile of lit cells, adapts automatically to changes in the product. That is precisely the defect: if the background level of the composite moves between release vintages, a percentile threshold moves with it, and the resulting margin absorbs the artefact instead of revealing it. A fixed threshold does the opposite: a shift in the background shows up in it as movement at the bottom of the radiance distribution, which is how the drift documented below becomes visible, and its level can be chosen so that the shift leaves the count of lit cells largely unaffected.

A \textbf{municipality-specific threshold}, for example a share of each unit's own baseline maximum, produces a margin that is not comparable across the cross-section, since the same physical brightness would count as lit in one municipality and dark in its neighbour. In a difference-in-differences design the comparison between treated and control units would then rest partly on the definition rather than on the data.

Within the fixed-threshold family the level of $\tau$ trades two errors against each other. Setting it too low leaves the margin exposed to the product's own treatment of the background, which does not hold still over the panel. \citet{falchetta2019high}, who clean annual medians of monthly composites with a floor of 0.25 nW/cm$^2$/sr, raise it to 0.35 nW from 2017, because pixels that should read zero, such as those of large water bodies, turn systematically positive from that year. The shift has a documented source in the processing of the sensor data: from 2017 the zero point of the day-night band no longer includes airglow, so that areas without artificial light record positive radiance, and \citet{coesfeld2020reducing} subtract 0.15 nW/cm$^2$/sr from the later data when they compare the two periods. The same break appears on our grid. Between 2016 and 2017 the number of Italian cells above 0.5 nW/cm$^2$/sr rises by 29.4 percent, the number above 1 nW by 17.0 percent and the number above 2 nW by 10.8 percent, a jump that fades as the threshold rises, as expected from a shift at the bottom of the radiance distribution.

The drift reaches the estimates twice. Directly, a low threshold saturates the units: once every cell of a municipality exceeds the threshold, its lit area cannot grow and the unit contributes nothing to the estimate, so that a treatment effect measured on that margin is attenuated or governed by whichever change in processing falls inside the window. Indirectly, and this channel reaches total radiance as well, a shift $\delta$ in the background adds roughly $\delta$ for every cell counted above a low threshold, an amount that weighs most, relative to the total, where artificial light is scarce; at national scale, shifts of this kind in the zero point can rival the sum of artificial light where lighting infrastructure is sparse \citep{kyba2017artificially,coesfeld2020reducing}. Baseline geography and demography predict how many near-threshold cells a municipality holds, so conditioning on them while measuring the outcome at a low threshold absorbs the artefact rather than a confounder. At the other end, setting $\tau$ too high discards the dim settlement lighting that is the object of interest in sparsely populated areas, and saturates the units in the opposite direction, since few cells ever qualify.

Table~\ref{tab:sat} reports, for thresholds between 0.25 and 2 nW/cm$^2$/sr, the share of Italian municipalities in which every cell already lies above the threshold. At 0.25 nW/cm$^2$/sr that share is already 72.9 percent in 2013 and reaches 100 percent from 2020 onward, so the margin is frozen and no estimate computed on it can be informative. At 0.5 nW the share climbs from 42.9 percent in 2013 to a peak of 74.8 percent in 2021, which is a strong and increasing ceiling. At 1 nW it never exceeds 36.2 percent between 2013 and 2022, leaving about two municipalities in three free to respond throughout. At 2 nW the ceiling is looser still, never above 13.8 percent, but the threshold now sits above much of the settlement lighting we wish to observe.

\begin{table}[htbp]
\centering
\caption{\textit{Share of municipalities with every cell above the threshold (\%)}}
\label{tab:sat}
\small
\begin{tabular}{lccccc}
\toprule
\textbf{Threshold (nW/cm$^2$/sr)} & \textbf{2013} & \textbf{2015} & \textbf{2017} & \textbf{2020} & \textbf{2022} \\
\midrule
0.25 & 72.9 & 61.2 & 98.8 & 100.0 & 100.0 \\
0.5 & 42.9 & 42.2 & 57.6 & 66.7 & 68.3 \\
1.0 (used here) & 20.1 & 21.6 & 28.7 & 33.5 & 33.6 \\
2.0 & 9.0 & 10.1 & 11.7 & 13.0 & 13.0 \\
\bottomrule
\end{tabular}
\begin{minipage}{0.92\linewidth}~\\
\justifying
\tiny
\noindent
Notes: all 7,903 Italian municipalities, unmasked \texttt{average} band of the VIIRS VNL annual composites, cells of $500 \times 500$ m in EPSG:32632. A municipality is counted as saturated when its lit-cell count equals its total cell count, so that its lit area cannot increase. The years shown are a selection of the estimation window; the saturated share peaks in 2021, at 74.8 percent for the 0.5 threshold, 36.2 percent for 1.0 and 13.8 percent for 2.0, and falls back in 2022. The non-monotone path at 0.25 nW between 2013 and 2017 and the 2021 peak at 0.5, 1 and 2 nW coincide with the movements of the background level reported in Table~\ref{tab:noise_floor}. \\
Sources: authors' elaboration based on \cite{elvidge2021annual}.
\end{minipage}
\end{table}

The second input to the choice is the level of the background. For global products the producers regard a single radiance threshold as impossible, because the background varies over space, rising at high latitudes, and over time \citep{elvidge2024statistical}. Italy lies between about 35 and 47 degrees north, inside the band where the average background of the annual product is lowest: in the annual composites background radiance lies overwhelmingly below 1 nW/cm$^2$/sr, and in 2015 it averaged under 0.2 nW between 50$^\circ$S and 50$^\circ$N \citep{elvidge2017viirs,elvidge2021annual}. The same group places the radiance that separates lit cells from background near 1 nW, applies a uniform cut of 2 nW in its own analysis, and distributes the unmasked grids so that users can set their own threshold \citep{elvidge2024statistical}. The value of 1 nW also defines lit pixels in the global VIIRS-like series of \citet{chen2021extended}, and fixed cut-offs between 0.3 and 5 nW are used to remove residual noise and to measure lit area across years \citep{kyba2017artificially,li2020harmonized,wang2022black}. The variation of the background over time is what Table~\ref{tab:noise_floor} measures on our own grid. The reference is the set of 93,438 cells assigned to an Italian municipality that have no built-up surface and no resident population within 2 km, where the recorded radiance is background. Their median radiance moves between 0.15 and 0.47 nW/cm$^2$/sr over 2013--2022 and jumps from 0.15 to 0.36 nW between 2016 and 2017, the year of the change in the airglow correction described above; in cells of open sea more than 20 km from the Italian coast the median moves from 0.03 to 0.24 nW between the same two years, as in the water bodies examined by \citet{falchetta2019high}. The share of these cells above 0.5 nW ranges between 2.5 and 36.9 percent, so that at 0.5 nW a count of lit cells partly records the treatment of the background. The share above 1 nW stays between 0.6 and 2.2 percent, and the 95th percentile never reaches 1 nW, peaking at 0.77 nW in 2021. In the estimation sample such cells account for at most 0.57 percent of the lit cells of funded municipalities and 0.30 percent of those of controls in any year, and the rate at which an unsettled cell exceeds 1 nW is similar in the two groups, between 0.3 and 1.1 percent.

\begin{table}[htbp]
\centering
\caption{\textit{Background radiance in unsettled cells and the 1 nW threshold (2013--2022)}}
\label{tab:noise_floor}
\footnotesize\setlength{\tabcolsep}{4pt}
\begin{tabular}{lcccccccc}
\toprule
 & \multicolumn{6}{c}{\textbf{Unsettled cells, whole country}} & \multicolumn{2}{c}{\textbf{Unsettled cells $\geq 1$ nW,}} \\
 & \multicolumn{3}{c}{Radiance, nW/cm$^2$/sr} & \multicolumn{3}{c}{Share above threshold, \%} & \multicolumn{2}{c}{\textbf{\% of lit cells}} \\
\cmidrule(lr){2-4}\cmidrule(lr){5-7}\cmidrule(lr){8-9}
Year & Median & P95 & P99 & $\geq 0.5$ & $\geq 1$ & $\geq 2$ & Funded & Controls \\
\midrule
2013 & 0.24 & 0.42 & 0.88 & 3.2 & 0.8 & 0.26 & 0.37 & 0.20 \\
2014 & 0.25 & 0.44 & 0.91 & 3.7 & 0.8 & 0.26 & 0.33 & 0.20 \\
2015 & 0.17 & 0.36 & 0.84 & 2.7 & 0.7 & 0.29 & 0.29 & 0.18 \\
2016 & 0.15 & 0.36 & 0.79 & 2.5 & 0.6 & 0.22 & 0.24 & 0.20 \\
2017 & 0.36 & 0.59 & 1.09 & 8.9 & 1.3 & 0.32 & 0.38 & 0.23 \\
2018 & 0.34 & 0.57 & 1.02 & 7.7 & 1.1 & 0.22 & 0.23 & 0.20 \\
2019 & 0.36 & 0.59 & 1.04 & 8.8 & 1.1 & 0.24 & 0.23 & 0.19 \\
2020 & 0.43 & 0.67 & 1.12 & 21.2 & 1.5 & 0.25 & 0.30 & 0.21 \\
2021 & 0.47 & 0.77 & 1.29 & 36.9 & 2.2 & 0.39 & 0.57 & 0.30 \\
2022 & 0.44 & 0.70 & 1.15 & 22.3 & 1.7 & 0.26 & 0.34 & 0.24 \\
\bottomrule
\end{tabular}
\begin{minipage}{0.98\linewidth}~\\
\justifying
\tiny
\noindent
Notes: unsettled cells are the 93,438 cells of $500 \times 500$ m assigned to an Italian municipality that have no built-up surface and no resident population within 2 km, according to the GHSL built-up and population grids for 2020 \citep{pesaresi2023ghsl,schiavina2023ghspop}; the 2 km buffer keeps light scattered from settlements out of the reference. Radiance is the \texttt{average} band of the annual VIIRS composite \citep{elvidge2021annual} before any truncation, on the same grid used to build the outcomes. Columns 2--7 describe the background level of the product in each year: its median moves between 0.15 and 0.47 nW/cm$^2$/sr across years, and the share of unsettled cells above 0.5 nW ranges between 2.5 and 36.9 percent, while the share above 1 nW stays between 0.6 and 2.2 percent and the 95th percentile never reaches 1 nW. The last two columns count the unsettled cells of funded and control municipalities that exceed 1 nW, as a share of all their cells above 1 nW, that is, the weight of cells that are lit without any settlement nearby in the lit-area outcome. \\
Sources: authors' elaboration based on \cite{elvidge2021annual,pesaresi2023ghsl,schiavina2023ghspop}.
\end{minipage}
\end{table}

\begin{figure}[htbp]
	\centering
	\caption{\textit{Event study on lit area at thresholds of 1, 2 and 5 nW/cm$^2$/sr (2011--2022)}}
	\includegraphics[width=0.95\linewidth]{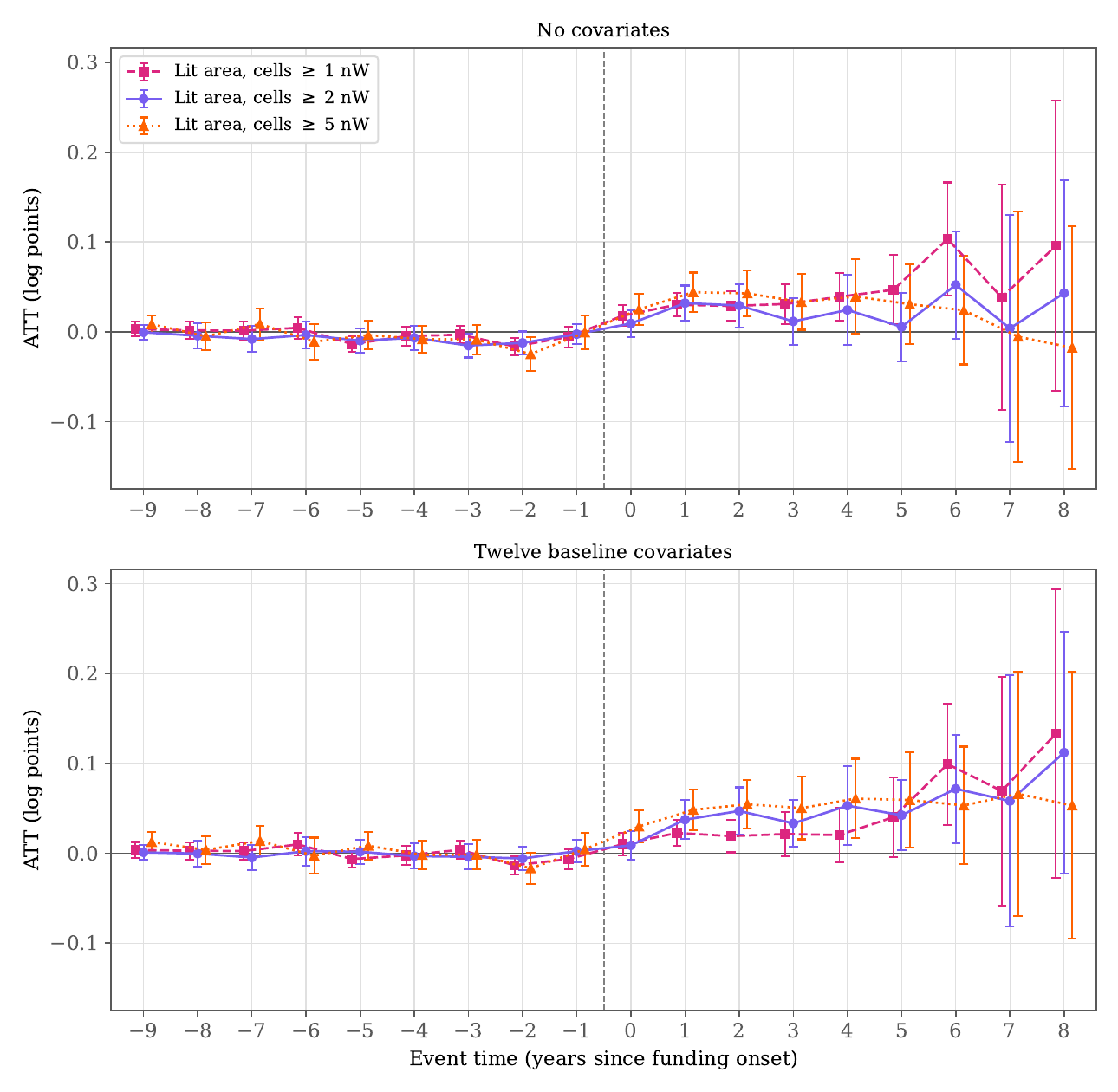}
	\label{fig:es_thresholds}
	\begin{minipage}{0.92\linewidth}
		\justifying
		\tiny
		\noindent
		Notes: each series re-estimates the design of Table~\ref{tab:main} on the log number of cells at or above the stated threshold, without covariates in the upper panel and with the twelve baseline (2013) covariates in the lower panel, 941 treated and 3,112 control municipalities, over every horizon observed in the window, from $e=-9$ to $e=+8$. Horizons $+7$ and $+8$ rest on 60 and 40 municipalities. Averages over post-onset horizons are reported in Table~\ref{tab:thresholds}. Bars are pointwise 95\% confidence intervals from a multiplier bootstrap with 300 replications, clustered at the municipality; the series are offset horizontally for legibility. \\
	\end{minipage}
\end{figure}

Raising the threshold moves the error to the other side. Even villages of a few hundred inhabitants in Europe and North America typically exceed 5 nW \citep{coesfeld2020reducing}, while on our grid, at 5 nW, 23.0 percent of the funded municipalities and 18.5 percent of the controls have no lit cell at all in 2013, and 26.7 and 20.7 percent in 2022, so that about one funded municipality in four has no lit area to measure. \citet{kyba2017artificially}, who measure lit area at this cut, note that lowering it to the smallest practical value would likely improve the correlation between changes in light and in output. Table~\ref{tab:thresholds} and Figure~\ref{fig:es_thresholds} repeat the estimation at 2 and 5 nW, and Section~\ref{section:results_discussion} discusses the comparison.

\begin{table}[htbp]
\centering
\caption{\textit{Post-onset ATT at alternative radiance thresholds (2011--2022)}}
\label{tab:thresholds}
\small
\begin{tabular}{lcccc}
\toprule
 & \multicolumn{2}{c}{\textit{All post-onset horizons}} & \multicolumn{2}{c}{\textit{Horizons 0 to +6}} \\
\cmidrule(lr){2-3} \cmidrule(lr){4-5}
 & \textbf{(1)} & \textbf{(2)} & \textbf{(3)} & \textbf{(4)} \\
 & \makecell{\textit{No}\\\textit{covariates}} & \makecell{\textit{Twelve}\\\textit{covariates}} & \makecell{\textit{No}\\\textit{covariates}} & \makecell{\textit{Twelve}\\\textit{covariates}} \\
\midrule
\multicolumn{5}{l}{\textit{Panel A. Threshold 1 nW/cm$^2$/sr}} \\[2pt]
Lit area (cells $\geq 1$ nW) & $+0.048^{***}$ & $+0.048^{**}$ & $+0.042^{***}$ & $+0.033^{***}$ \\
 & $(0.018)$ & $(0.020)$ & $(0.011)$ & $(0.012)$ \\[2pt]
Total radiance from lit area & $+0.037$ & $+0.041^{*}$ & $+0.046^{***}$ & $+0.036^{**}$ \\
 & $(0.024)$ & $(0.024)$ & $(0.013)$ & $(0.014)$ \\[2pt]
Radiance per lit cell & $-0.017$ & $-0.008$ & $-0.002$ & $+0.001$ \\
 & $(0.013)$ & $(0.013)$ & $(0.005)$ & $(0.007)$ \\[2pt]
\addlinespace[4pt]
\multicolumn{5}{l}{\textit{Panel B. Threshold 2 nW/cm$^2$/sr}} \\[2pt]
Lit area (cells $\geq 2$ nW) & $+0.023$ & $+0.051^{***}$ & $+0.023^{**}$ & $+0.042^{***}$ \\
 & $(0.016)$ & $(0.019)$ & $(0.011)$ & $(0.011)$ \\[2pt]
Total radiance from lit area & $+0.032$ & $+0.060^{*}$ & $+0.039^{***}$ & $+0.055^{***}$ \\
 & $(0.028)$ & $(0.032)$ & $(0.015)$ & $(0.017)$ \\[2pt]
Radiance per lit cell & $+0.002$ & $+0.005$ & $+0.014^{***}$ & $+0.013^{**}$ \\
 & $(0.011)$ & $(0.011)$ & $(0.005)$ & $(0.006)$ \\[2pt]
\addlinespace[4pt]
\multicolumn{5}{l}{\textit{Panel C. Threshold 5 nW/cm$^2$/sr}} \\[2pt]
Lit area (cells $\geq 5$ nW) & $+0.024$ & $+0.053^{***}$ & $+0.034^{**}$ & $+0.051^{***}$ \\
 & $(0.018)$ & $(0.019)$ & $(0.014)$ & $(0.015)$ \\[2pt]
Total radiance from lit area & $+0.026$ & $+0.078^{**}$ & $+0.038$ & $+0.069^{***}$ \\
 & $(0.038)$ & $(0.039)$ & $(0.024)$ & $(0.024)$ \\[2pt]
Radiance per lit cell & $+0.004$ & $+0.003$ & $+0.016^{**}$ & $+0.013^{*}$ \\
 & $(0.011)$ & $(0.013)$ & $(0.006)$ & $(0.007)$ \\[2pt]
\midrule
Treated municipalities & \multicolumn{4}{c}{941} \\
Control municipalities & \multicolumn{4}{c}{3,112} \\
\bottomrule
\end{tabular}
\begin{minipage}{0.92\linewidth}~\\
\justifying
\tiny
\noindent
Notes: each panel re-estimates the design of Table~\ref{tab:main}, changing only the radiance threshold above which a cell counts as lit: 1 nW/cm$^2$/sr in Panel A (the baseline), 2 and 5 nW/cm$^2$/sr in Panels B and C. Lit area is the log number of cells at or above the threshold, set to zero when no cell reaches it, total radiance the log of one plus the radiance summed over those cells, radiance per lit cell the log of their mean radiance. Columns (1) and (2) average the post-onset effects over every horizon observed in the window, columns (3) and (4) over horizons 0 to +6; columns (2) and (4) enter the twelve baseline (2013) covariates in the doubly-robust adjustment. Radiance per lit cell is defined only for municipalities with at least one lit cell in every year, which leaves 923 treated and 3,070 control municipalities at 1 nW, 874 and 2,921 at 2 nW, and 612 and 2,276 at 5 nW. The share of the 7,903 Italian municipalities in which every cell is lit is 20.1\% in 2013 and 33.6\% in 2022 at 1 nW, 9.0\% and 13.0\% at 2 nW, and 3.2\% and 3.9\% at 5 nW. The share of inner municipalities with no lit cell, treated and controls, is 1.2\% and 0.8\% in 2013 at 1 nW, 3.6\% and 3.6\% at 2 nW, and 23.0\% and 18.5\% at 5 nW; in 2022 the same shares are 0.4\% and 0.4\%, 4.1\% and 3.7\%, and 26.7\% and 20.7\%. Multiplier bootstrap standard errors with 300 replications, clustered at the municipality and computed with a fixed random seed, in parentheses. $^{***}$ $p<0.01$, $^{**}$ $p<0.05$, $^{*}$ $p<0.10$. \\
\end{minipage}
\end{table}

We therefore set $\tau = 1$ nW/cm$^2$/sr. In cells with no settlement nearby the value lies above the 95th percentile of background radiance in every year from 2013 to 2022 (Table~\ref{tab:noise_floor}), it leaves about two municipalities in three free to respond (Table~\ref{tab:sat}), and it lies below the brightness of ordinary settlement lighting, which preserves the signal.

The fixed threshold is a measurement-level answer to a problem that the satellite literature more often meets at the estimator level. A background level that does not hold still is a systematic error, and it is not independent of the characteristics on which the design conditions, since whether a cell clears a low threshold depends on how dim and how sparsely built the municipality is to begin with. \citet{alixgarcia2023remotely}, who face non-classical misclassification in a binary outcome read off satellite imagery, exploit knowledge of the remote-sensing process to model it inside the estimator. Here the error has a single known source, the position of the background level, so its reach can be limited in the measurement itself: the threshold is set once, in physical units, and every outcome is built above it, so that no part of the adjustment depends on the estimating equation or on the comparison group adopted downstream.

\subsubsection{Reconstructed opening years and sensor continuity}
\label{subsubsection:appendix_backcast}

The estimation window is 2011--2022, and its two opening years are reconstructed for a reason that lies in the product: the annual composite is derived from the monthly cloud-free composites of the day-night band, and the producer's monthly collection begins in April 2012. The first annual composite is accordingly distributed in two partial-year variants, one spanning April to December 2012 and one spanning April 2012 to March 2013; only the second carries the lit-area mask, and it overlaps the 2013 composite by a quarter.\footnote{Earth Observation Group, VIIRS Nighttime Light (VNL) product page: \url{https://eogdata.mines.edu/products/vnl/} [Accessed October 3, 2026].} Neither is a calendar year comparable with the rest of the series, so 2013 is the first complete calendar year the product covers, and the 2011 and 2012 values in the panel are an extrapolation. We keep them because they extend the pre-treatment path by two horizons, and we treat them as points that lie on a fitted line by construction, which add no independent evidence on pre-treatment parallelism, so that a diagnostic computed over them tests the extrapolation rather than the territory.

The 2011 and 2012 rows of the panel are produced by one rule, applied to each of the 7,903 Italian municipalities and stored with a source flag that separates them from observed rows. For each municipality, and separately for the lit-cell count and for the summed radiance above each threshold, a straight line is fitted by ordinary least squares to $\log(1+y)$ over the five observed years 2013--2017, extrapolated to 2011 and 2012, and transformed back, with negative predictions set to zero. Radiance per lit cell is not extrapolated on its own but recovered downstream as the ratio of the two extrapolated series, so that the identity $\log S = \log L + \log I$ holds in the reconstructed years as in the observed ones. The logarithm of one plus the value is used because both series can be zero: 45 municipalities have no lit cell at 1 nW/cm$^2$/sr in the reconstructed 2011, against 41 in the observed 2013. The fitting window matters, and we state how. It contains the step between 2016 and 2017 in which the national count of cells above 1 nW/cm$^2$/sr rises by 17.0 percent, from 449,929 to 526,557, a movement that Section~\ref{subsubsection:appendix_threshold} attributes to the product's treatment of the background rather than to the territory, and that step supplies the slope. The median municipal slope of the log lit-cell count is $+0.016$ per year when the line is fitted on 2013--2017 and zero to five decimal places when it is fitted on 2013--2016, so that the reconstructed national count of lit cells in 2011 lies below the observed 2013 count under the first window, at 430,164 against 457,601, and above it under the second, at 472,635. The sign of the pre-treatment slope that the two reconstructed years carry is thus a property of a processing choice, which is the systematic kind of error in the typology of \citet{jain2020benefits}, and the relevant question is how far it can reach.

Under the estimator of \citet{callaway2021difference} with not-yet-treated controls, every post-treatment group-time effect compares a post-onset year with the year before onset, and with a first cohort in 2014 the earliest such anchor is 2013, an observed year, so no reconstructed value enters a post-treatment comparison. We verified the implication on four versions of the panel that differ only in the two opening years: the extrapolation in use, fitted on 2013--2017; the same extrapolation refitted on 2013--2016, which excludes the step; a refit on 2013--2016 with the least-squares slope replaced by the Theil-Sen median of pairwise slopes; and no reconstruction at all, with the window opened in 2013. Table~\ref{tab:backcast} reports the four side by side. Across them, the post-onset average is unchanged to the tenth decimal place on each of the three outcomes and in each of the two specifications, at $+0.0478$ and $+0.0483$ on lit area, $+0.0366$ and $+0.0406$ on total radiance and $-0.0171$ and $-0.0082$ on radiance per lit cell, without and with the baseline covariates respectively, and the same identity holds between the 2011--2022 and the 2013--2022 windows when the outcomes are rebuilt on the median rather than the average summary of the composite. What the reconstruction does reach is the pre-treatment path, and it reaches further into it than the two opening horizons. With a varying base period each pre-treatment effect compares a year with the one before it, so the first lead of the 2014 cohort, at $e=-1$, compares the observed 2013 with the reconstructed 2012, and every horizon from $e=-9$ to $e=-1$ contains at least one cohort whose comparison involves a reconstructed year. The pre-treatment coefficients of the event study are therefore, in part, a statement about a fitted line, and we read them as such.

The alternative to extrapolating the two years is to take them from a series that measures them, and we examined both available families before settling on the extrapolation. The harmonised DMSP-VIIRS series of \citet{li2020harmonized} opens the window in 2011 on the digital-number scale of the older sensor, and the junction between its two sensors falls on the first treatment cohort and is not common to the two groups. Among the 3,879 inner municipalities of the harmonised panel that were not funded before 2017, for which the step from 2013 to 2014 lies entirely before treatment, the change in log light across that junction is $-0.507$ for the 768 first funded between 2017 and 2021 and $-0.370$ for the 3,111 never funded or funded from 2022 onward, all but one of the 3,112 controls of Table~\ref{tab:main}, a differential of $-0.137$ log points with $t=-6.4$ between two groups the policy had not yet reached. The VIIRS-like reconstruction of \citet{chen2021extended}, in the release that extends it to 1992--2024 \citep{chen2026npp}, expresses the DMSP era in radiance-like units and carries sensor discontinuities inside a series that looks continuous: brought onto our $500 \times 500$ m grid, its count of cells above 1 nW/cm$^2$/sr inside Italian municipal boundaries falls by 28.4 percent between 2010 and 2011, from 370,047 to 264,847, and rises by 45.9 percent between 2012 and 2013, from 264,699 to 386,312. A junction whose size differs between the units a design will treat and the units it will use as controls is error that is systematic and correlated with treatment, the case that \citet{jain2020benefits} distinguishes from random error uncorrelated with treatment and that \citet{wuepper2025satellite} list among the temporal inconsistencies produced by fusing data from different sensors, and in a difference-in-differences comparison it would be read as an effect of the policy. Neither series could have altered the post-onset estimates, for the reason given above, so all that a measured 2011 and 2012 could have added is pre-treatment evidence, and in both series that evidence would have been dominated by the discontinuity of the instrument.

In the terms introduced in Section~\ref{subsection:night_lights}, the reconstruction sits on the outcome side of the regression, its error is systematic, since its sign follows the fitting window, and the design confines its reach to the pre-treatment horizons. Two of the sources of temporal inconsistency listed by \citet{wuepper2025satellite} are at stake here: the fusion of data from different sensors in the harmonised and VIIRS-like series, and a change in processing in the 2016--2017 step that the fitting window contains. We therefore prefer an explicit extrapolation, whose assumption is visible and whose failure mode we have measured, to a splice that would import the discontinuity of the instrument. We keep the fitting window on which the panel was built, the first five observed years, and we do not choose among windows on the basis of the pre-treatment coefficients they produce.

\begin{table}[H]
\centering
\caption{\textit{The two reconstructed years under four versions of the panel}}
\label{tab:backcast}
\small
\resizebox{\textwidth}{!}{%
\begin{tabular}{lcccc}
\toprule
 & \multicolumn{2}{c}{\textbf{2011 and 2012}} & \multicolumn{2}{c}{\textbf{Lit area, post-onset ATT}} \\
\cmidrule(lr){2-3} \cmidrule(lr){4-5}
 & \textit{Trend} & \textit{Fit window} & \textit{No covariates} & \textit{Baseline covariates} \\
\midrule
Version in use & least squares & 2013--2017 & $+0.048^{***}$ & $+0.048^{**}$ \\
 & & & $(0.018)$ & $(0.020)$ \\[2pt]
Refit, shorter window & least squares & 2013--2016 & $+0.048^{***}$ & $+0.048^{**}$ \\
 & & & $(0.018)$ & $(0.020)$ \\[2pt]
Refit, median slope & Theil-Sen & 2013--2016 & $+0.048^{***}$ & $+0.048^{**}$ \\
 & & & $(0.018)$ & $(0.020)$ \\[2pt]
No reconstruction & none & none & $+0.048^{**}$ & $+0.048^{**}$ \\
 & & & $(0.020)$ & $(0.022)$ \\
\bottomrule
\end{tabular}%
}
\begin{minipage}{0.98\linewidth}~\\
\justifying
\tiny
\noindent
Notes: the four rows re-estimate the design of Table~\ref{tab:main}, 941 treated and 3,112 control municipalities, changing only the 2011 and 2012 values of the outcomes. Rows one to three reconstruct them by municipality-level extrapolation of $\log(1+y)$, with the stated estimator of the trend and on the stated window; row four drops them and opens the window in 2013, so that the event study runs from $e=-7$ rather than $e=-9$ to $e=+8$. The ATT is the post-onset average on log lit area at 1 nW/cm$^2$/sr; the point estimates coincide across the four rows, and with Table~\ref{tab:main}, to the tenth decimal place, and the same holds for total radiance and radiance per lit cell, not shown. Standard errors are computed with the same fixed random seed as Table~\ref{tab:main}, whose lit-area cells the first row reproduces; in row four, estimated on the shorter panel, they are slightly wider, and the estimate without covariates is significant at the 5 rather than the 1 percent level. Multiplier bootstrap standard errors with 300 replications, clustered at the municipality, in parentheses. $^{***}$ $p<0.01$, $^{**}$ $p<0.05$, $^{*}$ $p<0.10$. \\
\end{minipage}
\end{table}

\subsubsection{Construction of the outcomes}
\label{subsubsection:appendix_construction}

To define the outcomes, let $P_i$ be the set of cells whose centroid falls inside the boundary of municipality $i$, and $R_{pt}$ the radiance recorded in cell $p$ in year $t$, in nW/cm$^2$/sr. Fix a brightness threshold $\tau$, the same in every year, and call a cell \textit{lit} in year $t$ when $R_{pt} \geq \tau$. The indicator $\mathbf{1}\{R_{pt} \geq \tau\}$ acts as a switch, equal to one for a lit cell and to zero for a dark one, and the three outcomes follow from it.

\begin{itemize}\itemsep6pt
\item \textbf{Lit area}, refers to the surface of a territory that has been illuminated. Count the lit cells:
\begin{equation}
L_{it}(\tau) = \sum_{p \in P_i} \mathbf{1}\{R_{pt} \geq \tau\}
\label{eq:lit}
\end{equation}
Every lit cell adds one and every dark cell adds nothing, so $L$ is a count of cells.

\item \textbf{Total radiance from the lit area}, quantity of light emitted by the municipality. Sum radiance over the lit cells only:
\begin{equation}
S_{it}(\tau) = \sum_{p \in P_i} R_{pt}\,\mathbf{1}\{R_{pt} \geq \tau\}
\label{eq:total}
\end{equation}
Here the switch multiplies each cell's own brightness, so dark cells drop out and lit cells enter with the value the satellite recorded.

\item \textbf{Radiance per lit cell}, the brightness of the lit part on average. Divide one by the other:
\begin{equation}
I_{it}(\tau) = \frac{S_{it}(\tau)}{L_{it}(\tau)}
\label{eq:int}
\end{equation}
\end{itemize}

The three variables are tied together. Since $S = L \cdot I$ by construction, in logs $\log S_{it} = \log L_{it} + \log I_{it}$ whenever $L_{it} > 0$, and any estimator linear in the outcome inherits the decomposition: the effect on total light equals the effect on how much territory is lit plus the effect on how bright it is. This is what allows us to identify which margin moves.

The measure is computed from the rasters rather than from a pre-aggregated panel, because the threshold has to be applied cell by cell. Municipal boundaries are reprojected to the metric coordinate reference system of the raster, and all operations are performed projected. The boundaries are then rasterised once into an integer label grid aligned to the radiance grid, with the same transform and shape and with cells assigned by centroid, and the label grid is reused for every year sharing that grid. For each observed year and threshold, the indicator $\mathbf{1}\{R \geq \tau\}$ is formed on the flattened raster and accumulated per municipality with a single bin count, once with unit weights to obtain the lit area $L$ and once with weights $R$ to obtain the total radiance from the lit area $S$. The procedure is exact and requires one pass per raster. With cells of $500 \times 500$ m, lit area converts to 0.25 km$^2$ per cell where a physical unit is preferred to a count.  One property of the grid should be stated, because it governs how the counts are to be read. The VNL average band is produced on a 15 arc-second grid \citep{elvidge2021annual}, which is close to 500 m only in the north to south direction: the native cell measures about 463 m north to south, and east to west it runs from about 378 m at the southernmost latitude of the country to about 316 m at the northernmost, so our cells are coarser than the native cell in both directions everywhere in Italy. We request the band from Google Earth Engine \citep{gorelick2017google} at a scale of 500 m in EPSG:32632 without specifying a resampling method, so the platform's default applies and each target cell takes the value of the nearest native pixel rather than the average of the native pixels it overlaps. Reprojecting a raster changes its values through resampling, which makes the resampling rule part of the measurement rather than a technical detail \citep{wuepper2025satellite}. Between about three tenths and two fifths of the native pixels, depending on latitude, are therefore never read. The lit-cell count is a sample of the lit native pixels rather than a census of them, and the same holds for the sum of radiance over lit cells. Two features of the sampling bound what it can do to the estimates. The grid is the same in every year, so the set of native pixels a municipality contributes is fixed over the panel and enters its level, which the base period of the estimator differences away; it reaches the path only through which of the sampled pixels cross $\tau$ in a given year. And the rule that selects those pixels is the position of the grid, which is fixed by the projection and not by funding. What remains is noise in the year-to-year path of the outcome, the random error uncorrelated with treatment that \citet{jain2020benefits} separates from the systematic kind and that, in a dependent variable, increases standard errors rather than biasing the coefficient. The precision we report is therefore a lower bound on the precision the same design would deliver on the native grid. Applying the threshold on the native grid and counting native pixels directly would remove the sampling variance, and is the natural refinement of the construction.

\subsubsection{What the measure captures and its limitations}
\label{subsubsection:appendix_limits}

\begin{figure}[H]
\centering
\caption{\textit{New lights: cells crossing the 1 nW threshold between 2013 and 2022}}
\includegraphics[width=\linewidth]{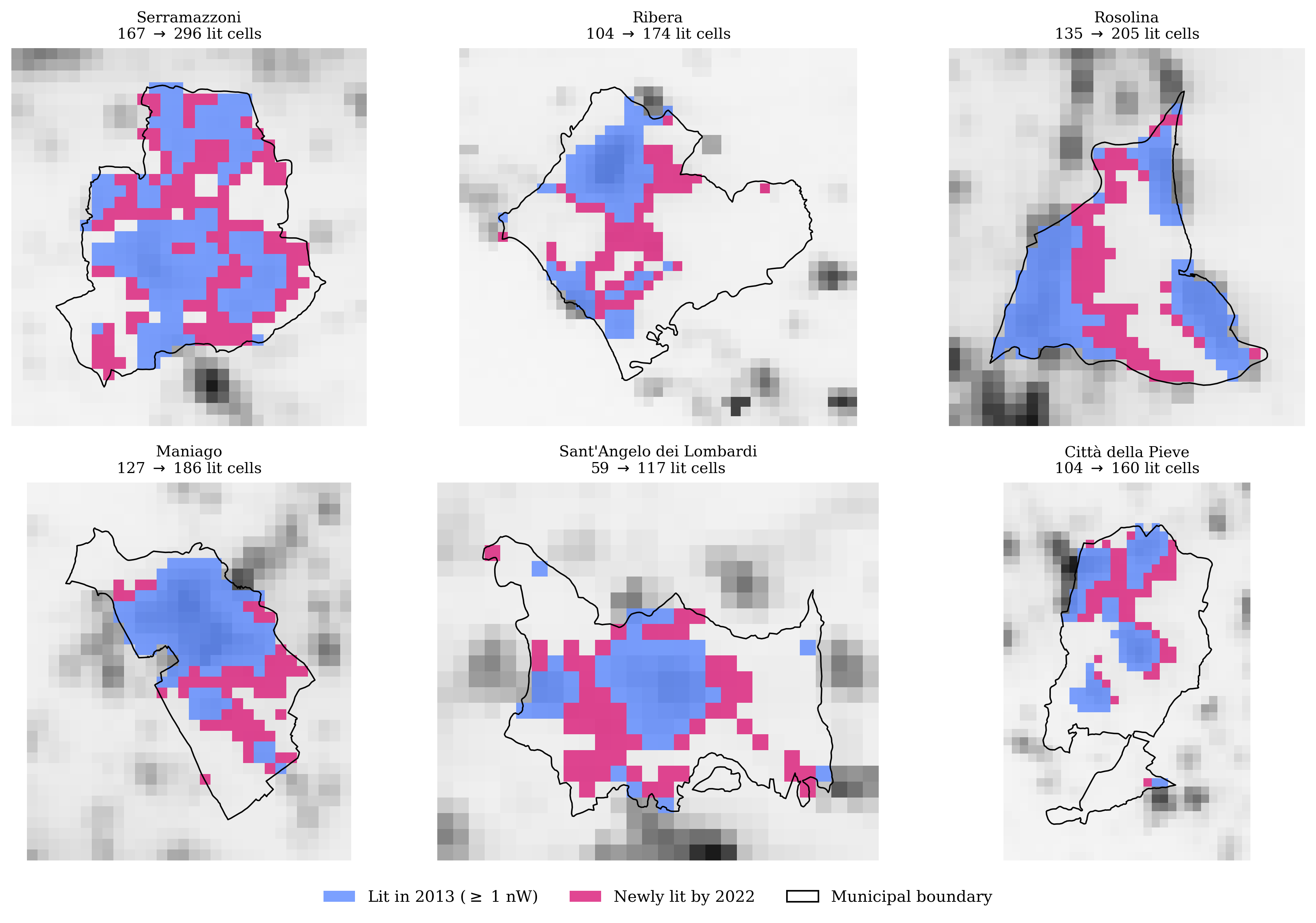}
\label{fig:insets}
\begin{minipage}{0.95\linewidth}
\justifying
\tiny
\noindent
Notes: six funded municipalities, one per region, selected among those with between 60 and 500 cells and the largest increase in lit cells, and shown for illustration only. Shading reports cells at or above 1 nW/cm$^2$/sr in 2013 and cells reaching that level only by 2022; the grey background is the 2022 radiance surface, and the outline is the municipal boundary. Cell counts in the panel titles refer to cells inside the boundary. \\
Sources: authors' elaboration based on \cite{elvidge2021annual,istat_confini,pcm_opencoesione}.
\end{minipage}
\end{figure}

The lit area records where light crosses a fixed physical brightness. Figure~\ref{fig:insets} makes the object of measurement concrete for six funded municipalities. Cells lit in 2013 form a compact core around the built-up centre, and the cells that cross the threshold by 2022 sit on its edge, along valley roads and around minor settlements. This is what the extensive margin records: an illuminated footprint that widens outward at its dim boundary, rather than a core that becomes brighter. The lit area is informative about the spatial extent of activity and infrastructure, and it is silent about the value of what is lit. A footprint expansion is not by itself evidence of income, output or welfare gains. This reading is consistent with the measurement literature: \cite{mellander2015nighttime} find that radiance tracks population density and establishment counts far more closely than wages or income, and \cite{bickenbach2016night} show that the cross-country lights-to-output elasticity does not transfer to units within a country.

The two margins carry different exposure to technological change, and the exposure is asymmetric. The transition of public lighting to blue-rich LED sources lowers measured radiance for the same physical lighting service, because a white LED emits part of its output at wavelengths below 500 nanometres, where the day-night band is effectively blind. \citet{kyba2017artificially} quantify the consequence for the very instrument that produces our outcome, reporting that a fall of roughly 30 percent in recorded radiance is compatible with a complete switch from high-pressure sodium lamps to LEDs at unchanged surface luminance, so that measured growth in radiance understates the growth of light in the human visual range. To the extent that the replacement was common to funded and unfunded municipalities, it is absorbed by the year effects of the estimator, and what matters is any differential in its pace between the two groups. If the strategy financed public-lighting works, that differential would depress measured radiance among the treated relative to the controls. The mechanism licenses one reading of our intensity results and not another. It means that a flat or negative intensive margin cannot be taken as evidence that no lighting investment took place, since an improvement in service and a fall in recorded radiance are compatible outcomes. It does not license attributing the aggregate intensity estimate to the channel, since that estimate, averaged over all post-treatment horizons, stands at $-0.017$ log points without covariates and at $-0.008$ with the baseline set, and an estimate that moves towards zero once baseline characteristics are absorbed behaves like the compositional imbalance documented for the all-municipality comparison rather than like a differential in lamp replacement, which characteristics measured in 2013 would not absorb. The extensive margin is exposed in the opposite and weaker sense, since the mechanism can only lower the radiance recorded for a given service and therefore cannot manufacture an upward crossing of $\tau$, and the same authors attribute the increases in radiance they observe elsewhere most plausibly to newly installed lighting. A threshold above the 95th percentile of the background radiance (Table~\ref{tab:noise_floor}) is unlikely to be crossed downward by the transition alone, although the mechanism can still push marginal cells below $\tau$.

One limitation is of a different order, because it concerns not what the measure does but what it has not been asked to do. The outcomes used here have never been compared with a measurement of the illuminated footprint taken on the ground, and this is a different point from the one made in the conclusions, where what is at stake is what light cannot record rather than how faithfully it records what it does. This matters beyond the usual appeal to caution, because the corrections for measurement error in remotely sensed variables that require no model of its source share one requirement. Whether the route is to model the relation between truth and measure and impute the unobserved truth, or to compute the estimate on the whole map and subtract the bias that a validation sample reveals, both need a calibration sample in which the satellite measure and an independent measure of the same quantity are observed together \citep{proctor2026parameter,angelopoulos2023prediction}. We do not have one, and in its absence the threshold of 1 nW/cm$^2$/sr rests on the documentation of the product and on the diagnostics reported above rather than on a comparison with ground measurements. A georeferenced register of municipal public-lighting points, or the municipal records of the lighting works themselves, observed for a subset of the municipalities of the panel over the years of the window and drawn as a probability sample of that panel rather than as a convenience set \citep{wuepper2025satellite}, would supply it. We state this as the principal open task of the measurement layer, and not as a residual caveat, because it is the step that would turn the diagnostics reported above into a quantified correction. Stating it that way also obliges us to say what such a register would have to contain, because in the settings where a comparison of this kind has actually been carried out the ground side of it turns out to be a measurement in its own right. \citet{lobell2020eyes} observe three ground measures of maize yield on the same smallholder plots in eastern Uganda, namely the farmer's own report of production and area, a crop cut on a sub-section of the plot, and a full-plot harvest, and they document large discrepancies among them, with the farmer reports that household surveys normally collect diverging sharply from the crop cuts. Judged against the full-plot harvests they treat as the gold standard, the satellite estimate accounts for at least as much of the variation as the cheaper ground measures do, so the instrument on the ground is not automatically the more reliable of the two. The condition this imposes on our own case is specific. A lighting register would discipline the threshold only in so far as it records the quantity the sensor sees, which is light emitted upward at the hour of overpass, rather than the fixtures a municipality has installed and declared, and a register that fails that condition would not settle the question but move it to the other side of the comparison.

Four further limitations are worth stating explicitly:
\begin{itemize}\itemsep2pt
\item \textbf{Ceiling}. Municipalities already saturated at $\tau$ cannot expand. In the estimation sample the ceiling binds almost only on the comparison group: at 1 nW the saturated share of controls rises from 4.6 percent in 2016 to 10.2 percent in 2017 and 16.1 percent in 2021, while among funded municipalities it never exceeds 0.7 percent. By curbing the measured growth of part of the comparison group, the ceiling can raise rather than lower the estimated effect on lit area. At 2 nW, where the saturated share of controls never exceeds 1.7 percent, the estimate with the baseline covariates is of similar size, $+0.051$ against $+0.048$ (Table~\ref{tab:thresholds});
\item \textbf{Granularity}. In small municipalities the lit-cell count is a small integer and moves in discrete jumps. Results are unchanged when the 185 municipalities below a 25-cell floor are excluded;
\item \textbf{Overglow}. Radiance spreads spatially beyond its source, so that a cell may cross $\tau$ because of a bright neighbour rather than because of local activity. A higher threshold, or restricting the cell set to built-up land, limits the concern;
\item \textbf{Boundary assignment}. Cells are assigned by centroid rather than by any overlap, which avoids double counting cells shared by adjacent municipalities.
\end{itemize}

\subsection{Estimation details}
\label{subsection:appendix_estimation}

As a benchmark, the difference-in-differences framework can be written as the standard two-way fixed effects (TWFE) specification:
\begin{equation}
Y_{i,t} = \eta_i + \lambda_t + \theta D_{i,t} + u_{i,t}
\label{eq.did_twfe}
\end{equation}
\noindent where $Y_{i,t}$ is one of the three outcomes in logs, $\eta_i$ captures municipality fixed effects, $\lambda_t$ time fixed effects, $D_{i,t}$ is the treatment indicator of Section~\ref{subsection:methodology} and $u_{i,t}$ is an idiosyncratic error term. In the presence of staggered treatment timing the TWFE estimator may produce biased estimates, as it combines already-treated and not-yet-treated units in the control group.

Two alternative comparison groups can be considered for $g \leq t$: (i) never-treated units ($C_i = 1$), and (ii) not-yet-treated units. The first leads to
\begin{equation}
ATT_{(g,t)} = \mathbb{E}[Y_{t}- Y_{g-1}\mid X, G = g] - \mathbb{E}[Y_{t} - Y_{g-1}\mid X, C = 1]
\label{eq.att_1}
\end{equation}
and the second to Equation~\ref{eq.att_2}, the comparison we use.

The estimator of \citet{callaway2021difference} requires covariates to be time-invariant. When time-varying covariates are included, their values are fixed at a reference (``base'') period. Specifically, for post-treatment periods, the base period corresponds to the last period before a given group becomes treated, while for pre-treatment periods it is the period immediately preceding the current one. This procedure ensures that covariates are not affected by treatment and preserves the validity of the identification strategy. Our covariates are measured in 2013, before the first cohort (for three municipalities, in the first year available), and held at that value in every year, so the choice of base period does not change them.

\subsection{Robustness to the annual summary of the composite}
\label{subsection:bands}

The lit area counts the cells whose annual radiance lies at or above a fixed threshold. At the dim edge of a settlement, where the extensive margin is measured, a cell can lie below the threshold in some months and above it in others, so that whether it counts as lit depends on how the annual composite summarises the year. The annual VNL product is built from the twelve monthly cloud-free average radiances of each cell and is distributed with four summaries of them: an average, which weights each month by its number of cloud-free observations, and the median, the maximum and the minimum of the monthly values. The producers found that the median approximates an annual average from which outliers have been removed, since it excludes high and low values such as those produced by seasonal biomass burning \citep{elvidge2021annual}. The maximum records the brightest month of the year and the minimum the darkest. Our outcomes are built on the average, without the background mask. We rebuilt the three outcomes on each of the other three summaries with the same zonal procedure, on the same $500 \times 500$ m grid and with the same threshold of 1 nW/cm$^2$/sr, reconstructed 2011 and 2012 on each of them as for the baseline outcomes, and re-estimated the design of Table~\ref{tab:main} with the same bootstrap seed, so that the rows for the average in Table~\ref{tab:bands} reproduce Table~\ref{tab:main} cell by cell.

At a common threshold, the four summaries light different sets of cells. In 2020 the maximum places 57.4 percent of the 1.2 million cells inside Italian municipal boundaries at or above 1 nW/cm$^2$/sr, the average 46.5 percent, the median 45.4 percent and the minimum 37.4 percent, as reported in the second column of Table~\ref{tab:bands}; in the 4,053 municipalities of the estimation sample, which are dimmer, the corresponding shares are 41.5, 29.5, 28.4 and 21.1 percent. The median is close to the average: across the 7,903 Italian municipalities the two measures of lit area correlate at 0.998 in logs in 2020, and 3,914 municipalities record the same number of lit cells on both. The maximum lights 132,051 cells that the average leaves dark, 10.9 percent of all Italian cells, while no cell is lit on the average and dark on the maximum. In those cells the median of the annual average is 0.84 nW/cm$^2$/sr and the median of the annual maximum is 1.15 nW/cm$^2$/sr: they lie close to the threshold, below it on average over the year and above it in the brightest month. The minimum counts a cell as lit only if none of its monthly values falls below the threshold. In 2020 it leaves 66 Italian municipalities without any lit cell, against 15 on the average, and at the median across municipalities it records a lit area 23.6 percent smaller than the average does. The minimum grid contains negative values in 1.8 percent of the cells in 2020, down to $-0.24$ nW/cm$^2$/sr, which we set to zero before applying the threshold; since the threshold lies far above zero, the counts are unaffected.

\begin{table}[htbp]
\centering
\caption{\textit{Post-onset ATT by annual summary of the VIIRS composite (2011--2022)}}
\label{tab:bands}
\small
\resizebox{\textwidth}{!}{%
\begin{tabular}{lccccccc}
\toprule
 & & \multicolumn{3}{c}{\textbf{No covariates}} & \multicolumn{3}{c}{\textbf{Twelve baseline covariates}} \\
\cmidrule(lr){3-5}\cmidrule(lr){6-8}
\textbf{Annual summary} & \makecell{\textbf{Cells $\geq 1$ nW}\\\textbf{in 2020 (\%)}} & \makecell{\textbf{Total radiance}\\\textbf{from lit area}} & \makecell{\textbf{Lit area}\\\textbf{($\geq 1$ nW)}} & \makecell{\textbf{Radiance per}\\\textbf{lit cell}} & \makecell{\textbf{Total radiance}\\\textbf{from lit area}} & \makecell{\textbf{Lit area}\\\textbf{($\geq 1$ nW)}} & \makecell{\textbf{Radiance per}\\\textbf{lit cell}} \\
\midrule
\multicolumn{8}{l}{\textit{Panel A. All post-onset horizons ($e=0,\dots,+8$)}} \\[2pt]
Average (baseline) & 46.5 & $+0.037$ & $+0.048^{***}$ & $-0.017$ & $+0.041^{*}$ & $+0.048^{**}$ & $-0.008$ \\
 & & $(0.024)$ & $(0.018)$ & $(0.013)$ & $(0.024)$ & $(0.020)$ & $(0.013)$ \\[2pt]
Median & 45.4 & $+0.040^{*}$ & $+0.052^{***}$ & $-0.016$ & $+0.044^{*}$ & $+0.050^{**}$ & $-0.005$ \\
 & & $(0.022)$ & $(0.019)$ & $(0.012)$ & $(0.023)$ & $(0.021)$ & $(0.012)$ \\[2pt]
Maximum & 57.4 & $+0.085^{***}$ & $+0.127^{***}$ & $-0.041^{***}$ & $+0.068^{***}$ & $+0.085^{***}$ & $-0.017$ \\
 & & $(0.025)$ & $(0.023)$ & $(0.013)$ & $(0.025)$ & $(0.021)$ & $(0.014)$ \\[2pt]
Minimum & 37.4 & $+0.004$ & $+0.011$ & $-0.023$ & $+0.041$ & $+0.043$ & $-0.010$ \\
 & & $(0.033)$ & $(0.026)$ & $(0.017)$ & $(0.035)$ & $(0.027)$ & $(0.016)$ \\[2pt]
\midrule
\multicolumn{8}{l}{\textit{Panel B. Horizons $e=0,\dots,+6$}} \\[2pt]
Average (baseline) &  & $+0.046^{***}$ & $+0.042^{***}$ & $-0.002$ & $+0.036^{**}$ & $+0.033^{***}$ & $+0.001$ \\
 & & $(0.013)$ & $(0.011)$ & $(0.005)$ & $(0.014)$ & $(0.012)$ & $(0.007)$ \\[2pt]
Median &  & $+0.046^{***}$ & $+0.045^{***}$ & $-0.002$ & $+0.038^{***}$ & $+0.036^{***}$ & $+0.002$ \\
 & & $(0.012)$ & $(0.010)$ & $(0.006)$ & $(0.014)$ & $(0.011)$ & $(0.006)$ \\[2pt]
Maximum &  & $+0.075^{***}$ & $+0.087^{***}$ & $-0.011$ & $+0.047^{***}$ & $+0.045^{***}$ & $+0.002$ \\
 & & $(0.011)$ & $(0.011)$ & $(0.008)$ & $(0.013)$ & $(0.012)$ & $(0.008)$ \\[2pt]
Minimum &  & $+0.028^{**}$ & $+0.026^{**}$ & $-0.004$ & $+0.045^{***}$ & $+0.043^{***}$ & $+0.001$ \\
 & & $(0.014)$ & $(0.011)$ & $(0.008)$ & $(0.016)$ & $(0.013)$ & $(0.008)$ \\[2pt]
\midrule
Treated municipalities & & 941 & 941 & 875--936 & 941 & 941 & 875--936 \\
Control municipalities & & 3,112 & 3,112 & 2,918--3,102 & 3,112 & 3,112 & 2,918--3,102 \\
\bottomrule
\end{tabular}%
}
\begin{minipage}{0.98\linewidth}~\\
\justifying
\tiny
\noindent
Notes: each row re-estimates the design of Table~\ref{tab:main}, changing only the annual summary of the VIIRS VNL composite from which the three outcomes are built. The annual composite is built from the twelve monthly cloud-free average radiances of each cell: the average weights each month by its number of cloud-free observations, and the median, the maximum and the minimum are taken over the monthly values \citep{elvidge2021annual}. All four grids are used without the background mask, on the same $500 \times 500$ m grid and with the same threshold of 1 nW/cm$^2$/sr; negative values of the minimum grid are set to zero before the threshold is applied. The second column reports the share of the 1,208,069 cells inside Italian municipal boundaries that lie at or above the threshold in 2020. Panel A averages the event-time effects over all post-onset horizons observed in the window, as in Table~\ref{tab:main}; Panel B averages them over horizons $0$ to $+6$, each of which is identified by at least three cohorts. On every summary 2011 and 2012 are reconstructed by municipality-level linear extrapolation in logs of the 2013--2017 path, as for the baseline outcomes. Radiance per lit cell is defined only for municipalities with at least one lit cell in every year, which leaves 923 and 3,070 treated and control municipalities on the average, 922 and 3,067 on the median, 936 and 3,102 on the maximum and 875 and 2,918 on the minimum. \citet{callaway2021difference} doubly-robust estimates, cohorts 2014--2021, comparison group never- and not-yet-treated eligible inner municipalities, no anticipation, window 2011--2022. All rows are estimated with the same bootstrap seed as Table~\ref{tab:main}, so that the rows for the average reproduce its estimates and standard errors. Bootstrap standard errors with 300 replications, clustered at the municipality, in parentheses. $^{***}$ $p<0.01$, $^{**}$ $p<0.05$, $^{*}$ $p<0.10$. \\
Sources: authors' elaboration based on \cite{elvidge2021annual}.
\end{minipage}
\end{table}

\begin{figure}[htbp]
    \centering
    \caption{\textit{Event study on lit area by annual summary of the VIIRS composite ($\geq 1$ nW/cm$^2$/sr), horizons up to $e=+6$ (2011--2022)}}
    \includegraphics[width=0.95\linewidth]{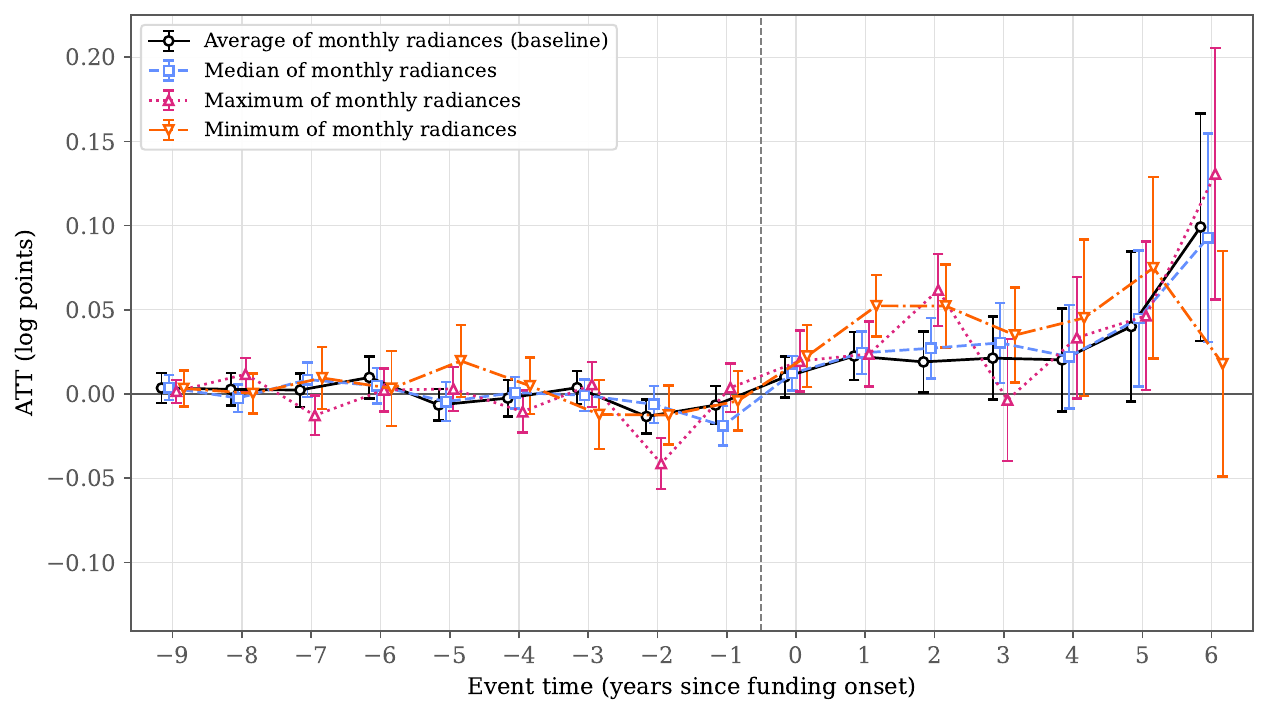}
    \label{fig:es_bands_lnlit1}
    \begin{minipage}{0.92\linewidth}
        \justifying
        \tiny
        \noindent
        Notes: \citet{callaway2021difference} doubly-robust estimates entering the twelve baseline (2013) covariates, cohorts 2014--2021, comparison group never- and not-yet-treated eligible inner municipalities, 941 treated and 3,112 control municipalities. Each series re-estimates the same design changing only the annual summary of the twelve monthly cloud-free radiances from which the lit area is built: the average, which is the baseline and coincides with the lit-area series of Figure~\ref{fig:es}, the median, the maximum and the minimum, on the same $500 \times 500$ m grid and with the same threshold of 1 nW/cm$^2$/sr. On every summary 2011 and 2012 are reconstructed as for the baseline outcomes. Over the nine pre-treatment horizons the largest coefficient in absolute value is $0.013$ on the average, $0.019$ on the median, $0.041$ on the maximum and $0.020$ on the minimum, and one, one, three and none of them respectively are individually significant at the 5 percent level. The horizontal axis stops at $e=+6$; averages over all post-onset horizons, including $e=+7$ and $e=+8$, are in Panel A of Table~\ref{tab:bands}. Bars are pointwise 95\% confidence intervals from a multiplier bootstrap with 300 replications, clustered at the municipality; the series are offset horizontally for legibility, and the dashed vertical line separates $e=-1$ from $e=0$. 
    \end{minipage}
\end{figure}

\begin{figure}[htbp]
	\centering
	\caption{\textit{Event study on radiance per lit cell by annual summary of the VIIRS composite ($\geq 1$ nW/cm$^2$/sr), horizons up to $e=+6$ (2011--2022)}}
	\includegraphics[width=0.95\linewidth]{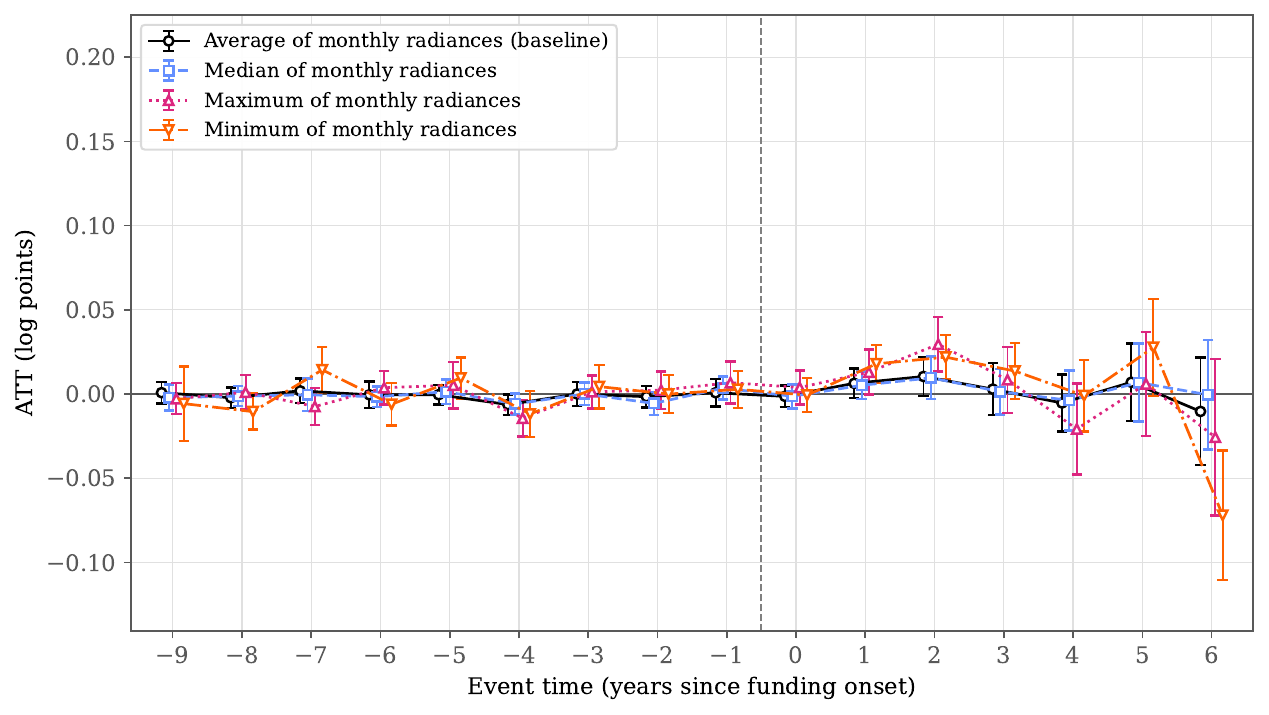}
	\label{fig:es_bands_lint1}
	\begin{minipage}{0.92\linewidth}
		\justifying
		\tiny
		\noindent
		Notes: same design and annual summaries as Figure~\ref{fig:es_bands_lnlit1}, for radiance per lit cell: \citet{callaway2021difference} doubly-robust estimates entering the twelve baseline (2013) covariates. The outcome is defined only for municipalities with at least one lit cell in every year, which leaves 923 treated and 3,070 control municipalities on the average, 922 and 3,067 on the median, 936 and 3,102 on the maximum and 875 and 2,918 on the minimum. Over the nine pre-treatment horizons the largest coefficient in absolute value is $0.006$ on the average, $0.006$ on the median, $0.014$ on the maximum and $0.014$ on the minimum, and one, none, one and one of them respectively are individually significant at the 5 percent level. The horizontal axis stops at $e=+6$; averages over post-onset horizons are in Table~\ref{tab:bands}. Bars are pointwise 95\% confidence intervals from a multiplier bootstrap with 300 replications, clustered at the municipality; the series are offset horizontally for legibility, and the dashed vertical line separates $e=-1$ from $e=0$. 
	\end{minipage}
\end{figure}

Table~\ref{tab:bands} reports the estimates over all post-onset horizons in Panel A and over horizons 0 to +6 in Panel B, and Figure~\ref{fig:es_bands_lnlit1} shows the event study of the lit area on the four summaries with the twelve baseline covariates; the corresponding figures for total radiance and for radiance per lit cell are in the Appendix \ref{subsection:appendix_ft} (Figures~\ref{fig:es_bands_lsum1} and~\ref{fig:es_bands_lint1}). On the median the estimates are close to the baseline. Over all post-onset horizons, the lit area expands by $+0.052$ log points without covariates and by $+0.050$ with them, against $+0.048$ and $+0.048$ on the average, both significant at the 5 percent level; radiance per lit cell stays at $-0.016$ and $-0.005$, and total radiance is not significant at the 5 percent level. In both panels and on every outcome, the estimates on the median and on the average differ by at most $0.004$ log points.

At the maximum, the estimates are larger. Over all post-onset horizons, the lit area expands by $+0.127$ log points without covariates and by $+0.085$ with them, and total radiance rises by $+0.085$ and $+0.068$, all four estimates significant at the 1 percent level, while radiance per lit cell moves by $-0.041$ without covariates, significant at the 1 percent level, and by $-0.017$ with them, not significant. The pattern is consistent with the construction of the summary, since the cells where the maximum adds lie close to the threshold, so that they raise the count of lit cells and lower the average radiance of the lit area. The maximum is also the summary on which the covariate adjustment moves the lit-area estimate most, from $+0.127$ to $+0.085$, and the one with the least flat pre-treatment path: with covariates, three of the nine pre-treatment coefficients on the lit area are individually significant at the 5 percent level, the largest being $-0.041$ at $e=-2$, against one on the average and one on the median.

On the minimum the estimates over all post-onset horizons are smaller or less precise. The lit area expands by $+0.011$ log points without covariates and by $+0.043$ with them, and neither estimate is significant at the 10 percent level: with covariates the point estimate is close to the $+0.048$ of the average but its standard error is larger, and without covariates it is pulled down by the two longest horizons, where the coefficients are $-0.067$ and $-0.012$ and rest on the 60 and 40 municipalities of the earliest cohorts. Over horizons 0 to +6 the lit-area estimate on the minimum is $+0.026$ without covariates and $+0.043$ with them, both significant at the 5 percent level, and no pre-treatment coefficient on the lit area is individually significant in either specification. Over all post-onset horizons total radiance and radiance per lit cell are not significant at the 5 percent level on the minimum, as on the average.

The expansion of the lit area is therefore not specific to the average summary of the composite. Its estimate is positive on every summary, in both panels and in both specifications, and over horizons 0 to +6 it is significant at the 5 percent level in all eight cases, between $+0.026$ and $+0.087$ log points; radiance per lit cell is negative or close to zero throughout, and total radiance carries a standard error at least as wide as that of the lit area in every column. The maximum returns the largest lit-area estimate in every column, as its definition leads one to expect, and given its less flat pre-treatment path we read it as an upper envelope of the extensive-margin response rather than as an estimate to prefer. We keep the average as the baseline, since the median reproduces its estimates, while the maximum admits cells that are lit only in the brightest month and the minimum only cells that are lit in every month.

\newpage

\subsection{Figures and tables}
\label{subsection:appendix_ft}

\begin{table}[htbp]
\centering
\caption{\textit{SNAI 2014--2020 projects by thematic areas}}
\label{tab:snai_projects_theme}
\resizebox{\textwidth}{!}{%
\begin{tabular}{@{}lllllllllll@{}}
\toprule
\multicolumn{1}{c}{\multirow{2}{*}{\textbf{Thematic area}}} & 
\multicolumn{5}{c}{\textbf{Projects}} & 
\multicolumn{5}{c}{\textbf{Financing}} \\
\cmidrule(l){2-11}
\multicolumn{1}{c}{} &
\multicolumn{1}{c}{\textit{\textbf{Number}}} &
\multicolumn{1}{c}{\textit{\textbf{Open}}} &
\multicolumn{1}{c}{\textit{\textbf{Liquidated}}} &
\multicolumn{1}{c}{\textit{\textbf{Not started}}} &
\multicolumn{1}{c}{\textit{\textbf{Ended}}} &
\multicolumn{1}{c}{\textit{\textbf{Total ({\euro})}}} &
\multicolumn{1}{c}{\textit{\textbf{Payments}}} &
\multicolumn{1}{c}{\textit{\textbf{EU}}} &
\multicolumn{1}{c}{\textit{\textbf{Public}}} &
\multicolumn{1}{c}{\textit{\textbf{Private}}} \\
\midrule
Administrative Capacity & 112 & 60.7\% & 9.8\% & 15.2\% & 14.3\% & 35,550,675 & 36.0\% & 21.5\% & 94.8\% & 5.2\% \\
Business Competitiveness & 1,107 & 28.5\% & 2.3\% & 1.1\% & 68.0\% & 100,394,089 & 74.6\% & 47.9\% & 80.7\% & 19.3\% \\
Culture and Tourism & 273 & 22.3\% & 5.1\% & 14.7\% & 57.9\% & 87,764,309 & 53.3\% & 41.3\% & 100.0\% & 0.0\% \\
Digital Networks and Services & 80 & 38.8\% & 7.5\% & 33.8\% & 20.0\% & 82,597,920 & 48.4\% & 49.8\% & 99.8\% & 0.2\% \\
Education and Training & 451 & 34.8\% & 4.4\% & 29.5\% & 31.3\% & 61,378,909 & 38.3\% & 9.5\% & 99.8\% & 0.2\% \\
Employment and Work & 326 & 6.4\% & 43.6\% & 2.1\% & 47.9\% & 10,161,079 & 68.4\% & 55.9\% & 99.4\% & 0.6\% \\
Energy & 142 & 29.6\% & 7.0\% & 4.2\% & 59.2\% & 44,134,756 & 59.6\% & 47.0\% & 99.3\% & 0.7\% \\
Environment & 295 & 44.4\% & 7.1\% & 21.0\% & 27.5\% & 70,155,140 & 44.5\% & 27.5\% & 100.0\% & 0.0\% \\
Research and Innovation & 102 & 19.6\% & 0.0\% & 1.0\% & 79.4\% & 19,124,301 & 73.1\% & 40.6\% & 79.5\% & 20.5\% \\
Social Inclusion and Health & 364 & 40.9\% & 6.9\% & 30.5\% & 21.7\% & 110,587,516 & 30.0\% & 10.6\% & 98.5\% & 1.5\% \\
Transport and Mobility & 313 & 42.5\% & 2.2\% & 21.1\% & 34.2\% & 169,259,755 & 46.9\% & 43.8\% & 99.8\% & 0.2\% \\
Total & 3,565 & 31.7\% & 7.9\% & 13.5\% & 46.9\% & 791,108,450 & 49.2\% & 35.2\% & 96.5\% & 3.5\% \\
\bottomrule
\end{tabular}%
}\\[0.3em]
\begin{minipage}{0.98\linewidth}
\justifying
\tiny
\noindent
Notes: data updated as of October 31, 2025. \\
Sources: authors' elaboration based on \cite{pcm_opencoesione}.
\end{minipage}
\end{table}


\begin{figure}[htbp]
	\centering
	\caption{\textit{Funding allocated to Italian municipalities under SNAI 2014--2020}}
	\includegraphics[width=0.8\linewidth]{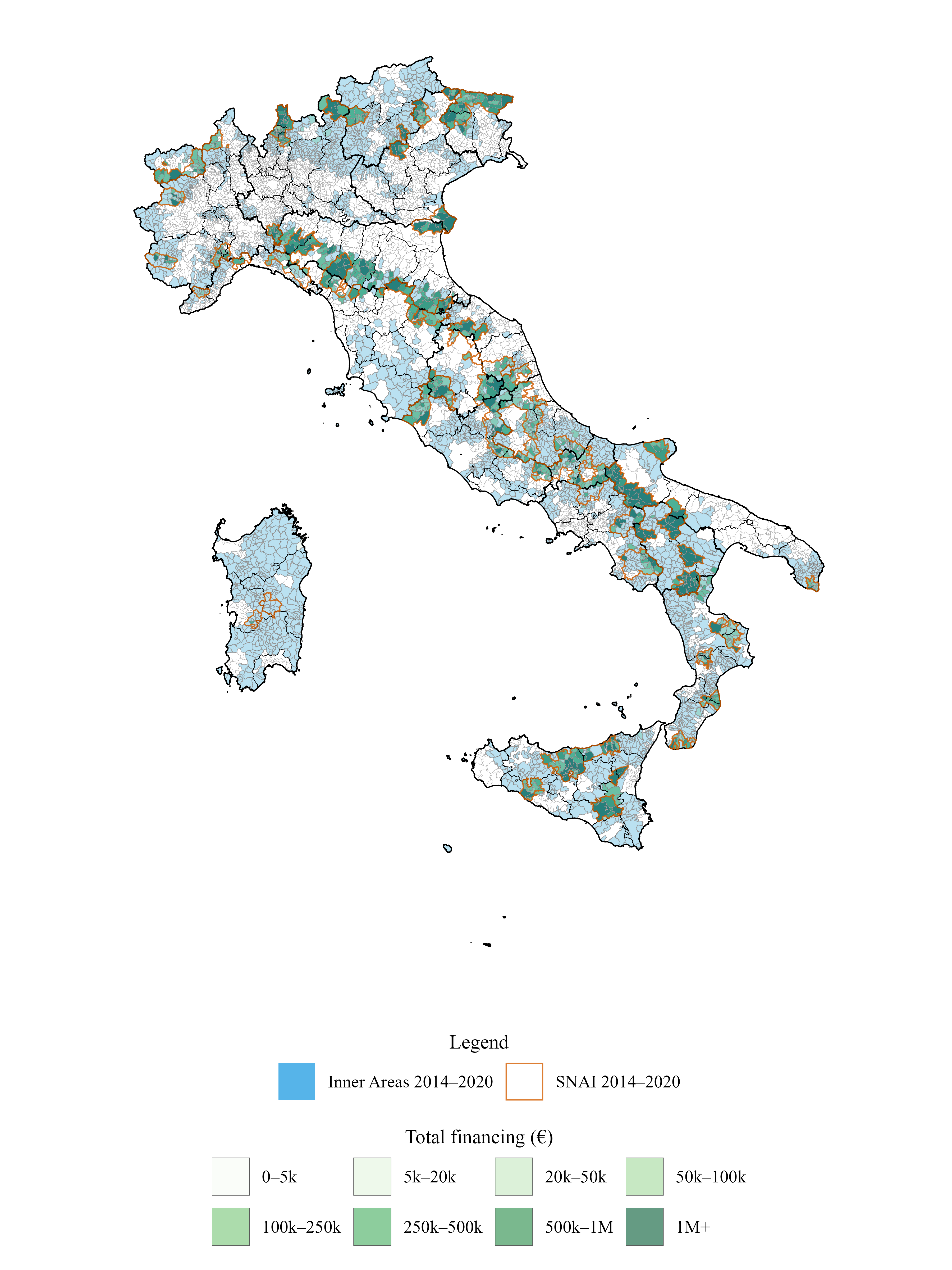}
	\label{fig:map_funds}
	\begin{minipage}{0.8\linewidth}
		\justifying
		\tiny
		\noindent
		Notes: financed municipalities for the 2014--2020 SNAI have been retrieved from \href{https://opencoesione.gov.it/it/dati/}{OpenCoesione} \citep{pcm_opencoesione}, data updated as of October 31, 2025. We considered as funded all municipalities with at least one active project (even if not yet started). Within the category `Inner Areas 2014--2020' we include municipalities classified as intermediate, peripheral, and ultra-peripheral. \\
		Sources: authors' elaboration based on \cite{istat_aree_interne,istat_confini} and \cite{pcm_aree_snai,pcm_opencoesione}.
	\end{minipage}
\end{figure}


\begin{figure}[H]
	\centering
	\caption{\textit{Treatment timing across Inner Area municipalities}}
	\label{fig:map_treatment}

	\noindent
	\begin{minipage}{0.49\linewidth}
		\centering
		\textbf{a}
	\end{minipage}
	\hfill
	\begin{minipage}{0.49\linewidth}
		\centering
		\textbf{b}
	\end{minipage}

	\vspace{0.1cm}

	\noindent
	\begin{minipage}{0.49\linewidth}
		\centering
		\includegraphics[width=\linewidth]{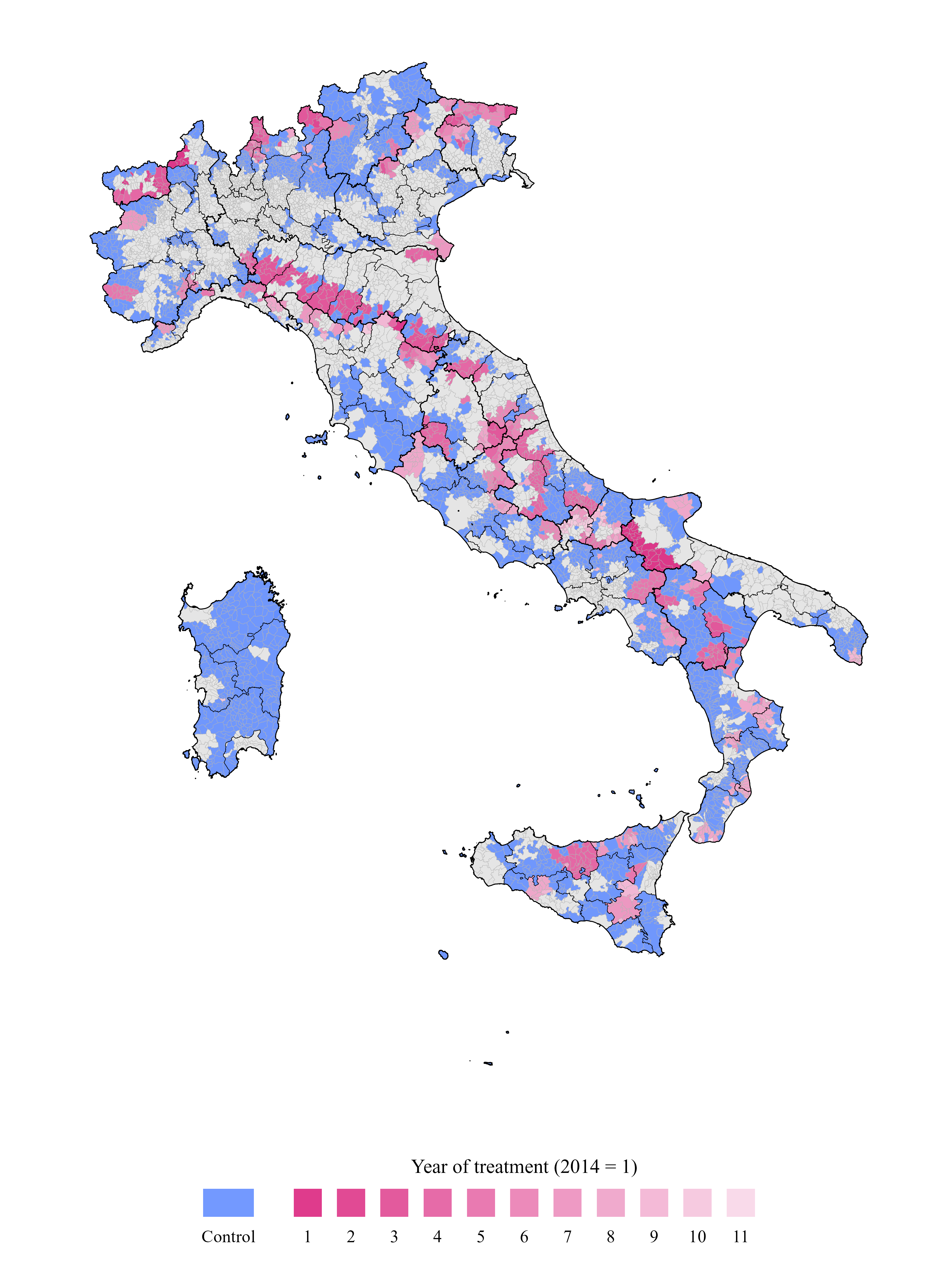}
	\end{minipage}
	\hfill
	\begin{minipage}{0.49\linewidth}
		\centering
		\includegraphics[width=\linewidth]{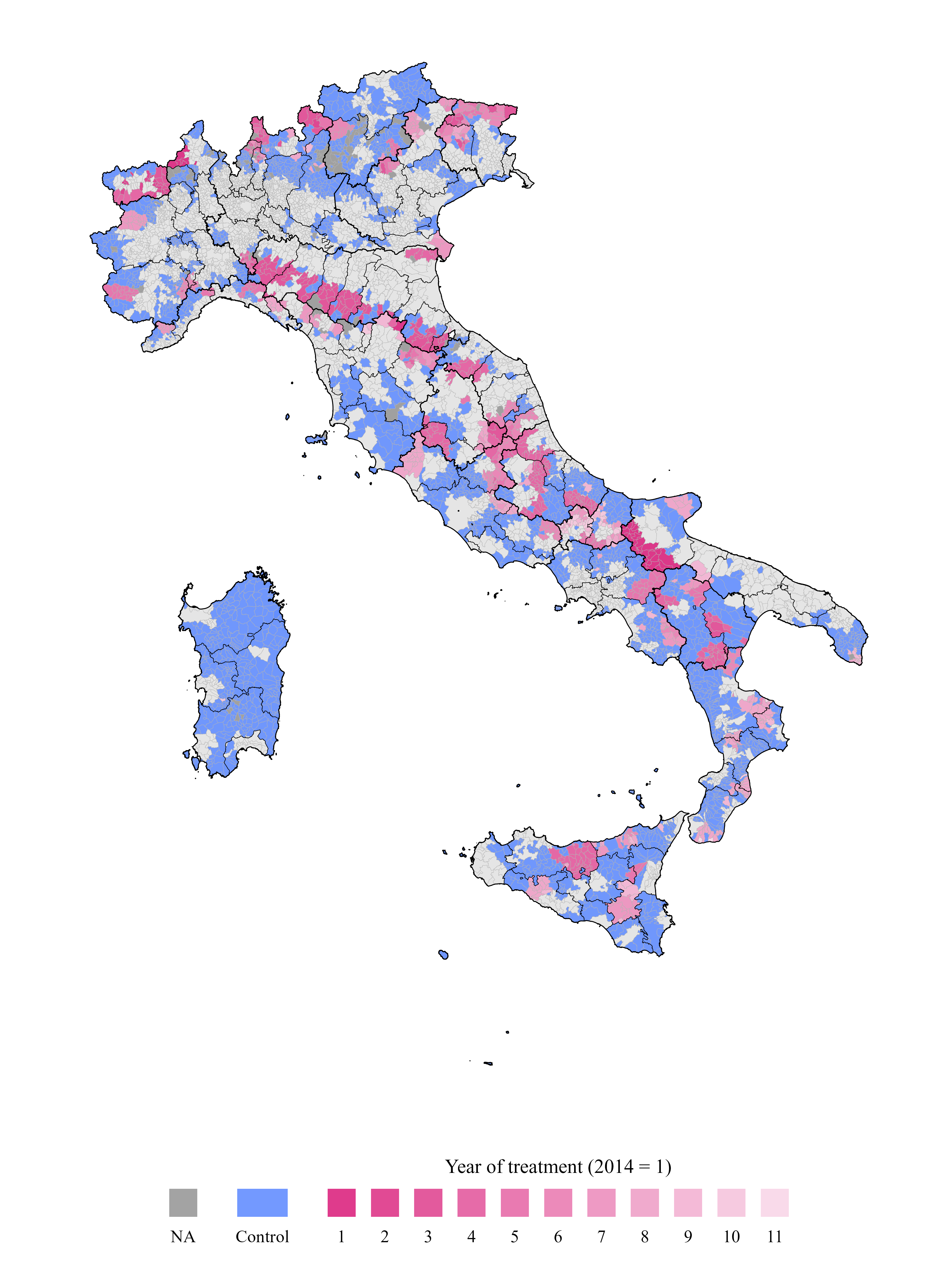}
	\end{minipage}

	\vspace{0.2cm}

	\begin{minipage}{0.95\linewidth}
		\justifying
		\tiny
		\noindent
		Notes: panel A shows the treatment-count classification for Inner Area municipalities. Panel B additionally identifies municipalities excluded from the analysis because of missing values. \\
        Sources: authors' elaboration based on \cite{istat_aree_interne,istat_confini} and \cite{pcm_aree_snai,pcm_opencoesione}.
	\end{minipage}
\end{figure}

\begin{table}[htbp]
\centering
\caption{\textit{Treated units among inner areas municipalities (2011--2022)}}
\label{tab:treated}
\resizebox{0.85\columnwidth}{!}{%
\begin{tabular}{@{}rrrrrr@{}}
\toprule
\multicolumn{1}{c}{\multirow{2}{*}{\textbf{Year}}} & \multicolumn{2}{c}{\textbf{Municipalities}} & \multicolumn{3}{c}{\textbf{Funds ({\euro})}} \\
\cmidrule(lr){2-6} 
\multicolumn{1}{c}{} & \multicolumn{1}{c}{\textit{\textbf{Treated}}} & \multicolumn{1}{c}{\textit{\textbf{Share (\%)}}} & \multicolumn{1}{c}{\textit{\textbf{Total}}} & \multicolumn{1}{c}{\textit{\textbf{Commercial}}} & \multicolumn{1}{c}{\textit{\textbf{Housing}}} \\ \midrule
2011 & 0 & 0.0\% & 0 & 0 & 0 \\
2012 & 0 & 0.0\% & 0 & 0 & 0 \\
2013 & 0 & 0.0\% & 0 & 0 & 0 \\
2014 & 40 & 1.0\% & 29,698,688 & 395,000 & 8,984,634 \\
2015 & 60 & 1.5\% & 30,484,335 & 395,000 & 9,545,602 \\
2016 & 173 & 4.3\% & 84,060,845 & 1,458,958 & 16,197,961 \\
2017 & 393 & 9.7\% & 158,004,218 & 44,051,821 & 26,672,593 \\
2018 & 504 & 12.4\% & 228,633,865 & 61,892,282 & 48,315,727 \\
2019 & 664 & 16.4\% & 323,272,929 & 85,422,531 & 63,663,803 \\
2020 & 789 & 19.5\% & 412,732,214 & 113,256,056 & 78,499,158 \\
2021 & 941 & 23.2\% & 528,220,352 & 139,184,296 & 105,200,108 \\
2022 & 992 & 24.5\% & 600,319,852 & 146,265,223 & 129,214,126 \\
\bottomrule
\end{tabular}%
}
\begin{minipage}{11cm}~\\
\justifying
\tiny 
Notes: OpenCoesione records projects mainly at the municipal and project-area levels and more rarely at the regional level. A project recorded above the municipal level is assumed to reach every targeted municipality within that level, and its amount is allocated across them in proportion to population. The cumulative count reaches 992 in 2022 because it includes the 51 municipalities first funded in that year, the last of the window, which would contribute only their onset year as a post-onset observation and are kept as controls instead; with the 36 first funded in 2023 and the 5 first funded in 2024, they form the 92 late-funded controls of Section~\ref{subsection:data}. \\
Source: Authors' elaboration based on \cite{pcm_opencoesione}.
\end{minipage}
\end{table}


\begin{table}[htbp]
\centering
\caption{\textit{Descriptive statistics}}
\label{tab:desc}
\resizebox{\textwidth}{!}{%
\begin{tabular}{l ccccc ccccc c}
\toprule
 & \multicolumn{5}{c}{\textbf{Treated} (941)} & \multicolumn{5}{c}{\textbf{Controls} (3,112)} & \\
\cmidrule(lr){2-6}\cmidrule(lr){7-11}
\textbf{Variable} & \textit{Mean} & \textit{SD} & \textit{P25} & \textit{Median} & \textit{P75} & \textit{Mean} & \textit{SD} & \textit{P25} & \textit{Median} & \textit{P75} & \makecell{\textbf{Norm.}\\\textbf{diff.}} \\
\midrule
\multicolumn{12}{l}{\textit{Night-light outcomes (2013)}} \\[2pt]
Lit area (cells $\geq 1$ nW/cm$^2$/sr) & 32.9 & 32.7 & 13.0 & 23.0 & 40.0 & 42.7 & 51.3 & 14.0 & 27.0 & 51.0 & $-0.23$ \\
Total radiance from lit area & 130.5 & 179.0 & 36.1 & 75.5 & 157.7 & 212.5 & 379.9 & 43.9 & 98.8 & 229.6 & $-0.28$ \\
Radiance per lit cell & 3.57 & 1.79 & 2.33 & 3.10 & 4.45 & 4.14 & 2.35 & 2.59 & 3.60 & 5.03 & $-0.27$ \\
Municipal cells ($500$ m grid) & 210 & 187 & 84 & 155 & 272 & 167 & 187 & 57 & 104 & 200 & $+0.23$ \\
\multicolumn{12}{l}{\textit{Geography}} \\[2pt]
Surface (km$^2$) & 52.5 & 46.9 & 20.5 & 39.3 & 68.2 & 41.8 & 46.7 & 14.4 & 26.2 & 49.9 & $+0.23$ \\
Altitude (m) & 628 & 294 & 423 & 612 & 794 & 447 & 313 & 215 & 405 & 630 & $+0.60$ \\
Coastline (km) & 0.71 & 4.98 & 0.00 & 0.00 & 0.00 & 1.23 & 5.88 & 0.00 & 0.00 & 0.00 & $-0.10$ \\
\multicolumn{12}{l}{\textit{Demography (2013)}} \\[2pt]
Resident population & 2,167 & 3,044 & 578 & 1,262 & 2,671 & 3,615 & 6,000 & 800 & 1,736 & 3,888 & $-0.30$ \\
Female population (\%) & 50.6 & 2.0 & 49.7 & 50.9 & 51.8 & 50.6 & 1.8 & 49.8 & 50.7 & 51.6 & $+0.02$ \\
Foreign population (\%) & 5.4 & 4.0 & 2.4 & 4.5 & 7.5 & 5.8 & 4.5 & 2.2 & 4.4 & 8.3 & $-0.09$ \\
Population over 65 (\%) & 26.6 & 6.9 & 22.0 & 25.6 & 29.7 & 22.7 & 5.6 & 18.9 & 22.0 & 25.7 & $+0.62$ \\
Birth rate & 6.3 & 3.4 & 4.4 & 6.3 & 8.3 & 7.3 & 3.1 & 5.4 & 7.4 & 9.1 & $-0.30$ \\
Mortality rate & 15.4 & 6.9 & 11.1 & 14.3 & 18.7 & 12.4 & 5.4 & 9.0 & 11.4 & 14.6 & $+0.50$ \\
Net migration rate & -1.4 & 32.6 & -16.1 & -0.9 & 15.0 & -0.4 & 30.0 & -13.7 & -1.8 & 12.4 & $-0.03$ \\
\multicolumn{12}{l}{\textit{Institutional quality (MAQI)}} \\[2pt]
Bureaucratic capacity & 102.4 & 3.8 & 100.6 & 102.5 & 104.3 & 101.7 & 3.7 & 100.3 & 102.0 & 103.6 & $+0.18$ \\
Quality of local politicians & 100.9 & 8.3 & 94.8 & 100.5 & 106.2 & 102.0 & 8.3 & 96.3 & 102.0 & 107.1 & $-0.13$ \\
Fiscal and economic performance & 98.5 & 5.8 & 95.3 & 99.3 & 102.3 & 98.8 & 5.8 & 95.8 & 99.9 & 102.6 & $-0.05$ \\\midrule
Municipalities & \multicolumn{5}{c}{941} & \multicolumn{5}{c}{3,112} & \\
\bottomrule
\end{tabular}%
}
\begin{minipage}{\textwidth}~\\
\justifying
\tiny
Notes: estimation sample of IA municipalities, 941 treated (first funded 2014--2021) and 3,112 never- or not-yet-funded controls. Night-light outcomes are measured in 2013 above the 1 nW/cm$^2$/sr threshold; lit-area cells convert to km$^2$ at 0.25 km$^2$ per cell. Demographic and institutional covariates are measured in 2013; the institutional-quality indices are imputed at the 2013 provincial mean for the 97 municipalities created by post-2013 mergers. The normalized difference is the difference in means divided by the square root of the average of the two group variances. Variable codes: surface (\texttt{sur}), altitude (\texttt{alt}), coastline (\texttt{coa}), resident population (\texttt{pop}), female population (\texttt{pop\_fem\_p}), foreign population (\texttt{pop\_for\_p}), population over 65 (\texttt{pop\_ov65\_p}), birth rate (\texttt{bir\_rat}), mortality rate (\texttt{mor\_rat}), net migration rate (\texttt{tot\_mig\_rat}), bureaucratic capacity (\texttt{maqi\_p1\_bur}), quality of local politicians (\texttt{maqi\_p2\_pol}), fiscal and economic performance (\texttt{maqi\_p3\_eco}). \\
Source: Authors' elaboration based on \cite{uval2014strategia,cerqua2025municipal,istat_aree_interne,istat_amdc,istat_confini,istat_geografia,istat_demo,pcm_opencoesione}.
\end{minipage}
\end{table}

\begin{figure}[htbp]
	\centering
	\caption{\textit{Event study on total radiance from lit area by annual summary of the VIIRS composite ($\geq 1$ nW/cm$^2$/sr), horizons up to $e=+6$ (2011--2022)}}
	\includegraphics[width=0.95\linewidth]{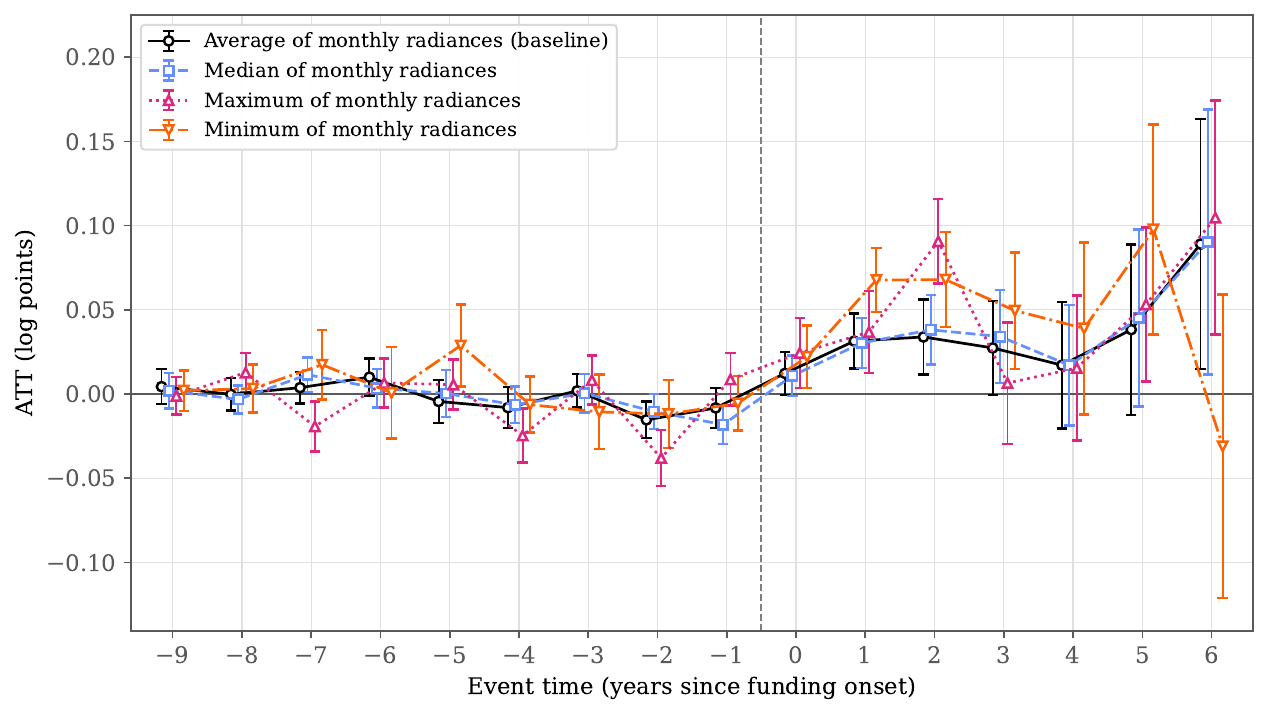}
	\label{fig:es_bands_lsum1}
	\begin{minipage}{0.92\linewidth}
		\justifying
		\tiny
		\noindent
		Notes: same design and annual summaries as Figure~\ref{fig:es_bands_lnlit1}, for total radiance from the lit area: \citet{callaway2021difference} doubly-robust estimates entering the twelve baseline (2013) covariates, 941 treated and 3,112 control municipalities. Over the nine pre-treatment horizons the largest coefficient in absolute value is $0.015$ on the average, $0.018$ on the median, $0.038$ on the maximum and $0.029$ on the minimum, and one, two, four and one of them respectively are individually significant at the 5 percent level. The horizontal axis stops at $e=+6$; averages over post-onset horizons are in Table~\ref{tab:bands}. Bars are pointwise 95\% confidence intervals from a multiplier bootstrap with 300 replications, clustered at the municipality; the series are offset horizontally for legibility, and the dashed vertical line separates $e=-1$ from $e=0$. 
	\end{minipage}
\end{figure}

\end{document}